\documentclass[aps,prl,twocolumn,superscriptaddress]{revtex4-1}
\usepackage[T1]{fontenc}
\usepackage[utf8]{inputenc}
\usepackage{amssymb}
\usepackage{amsmath}
\usepackage{color}
\usepackage{graphicx}
\usepackage{footmisc,booktabs}

\begin{document}

\title{SmO thin films: a flexible route to correlated flat bands with nontrivial topology}

\author{Deepa Kasinathan}
\affiliation{Max-Planck-Institut f\"{u}r Chemische Physik fester Stoffe\\N\"{o}thnitzer Str. 40, 01187 Dresden, Germany}

\author{Klaus Koepernik}
\affiliation{IFW Dresden, P.O. Box 270116, D-01171 Dresden, Germany}

\author{L. H. Tjeng}
\affiliation{Max-Planck-Institut f\"{u}r Chemische Physik fester Stoffe\\N\"{o}thnitzer Str. 40, 01187 Dresden, Germany}

\author{Maurits W. Haverkort}
\affiliation{Max-Planck-Institut f\"{u}r Chemische Physik fester Stoffe\\N\"{o}thnitzer Str. 40, 01187 Dresden, Germany}


\begin{abstract}
Using density functional theory based calculations, we show that the correlated mixed-valent compound
SmO is a 3D strongly topological semi-metal as a result of a 4$f$-5$d$ band inversion at the X point.
The [001] surface Bloch spectral density reveals two weakly interacting Dirac cones that are quasi-degenerate at the $\bar{M}$-point
and another single Dirac cone at the $\bar{\Gamma}$-point.
We also show that the topological non-triviality in SmO is very robust and prevails for a wide range
of lattice parameters, making it an ideal candidate to investigate topological nontrivial correlated
flat bands in thin-film form. Moreover, the electron filling is tunable by strain. 
In addition, we find conditions for which the inversion is of the 4$f$-6$s$ type,
making SmO to be a rather unique system.
The similarities of the crystal symmetry and the lattice
constant of SmO to the well studied ferromagnetic semiconductor EuO, makes SmO/EuO thin film interfaces
 an excellent contender towards realizing the quantum anomalous Hall effect in a strongly
 correlated electron system. 

\end{abstract}

\pacs{??}

\maketitle

Topological insulators (TI) are materials which exhibit a fundamentally new physical phenomena that was first predicted by theorists\,\cite{Kane_2005,Kane_2005_2,Fu_2007,Moore_2007,Roy_2009,Fu_2007_2,Zhang_2009} and
subsequently ascertained in 
experiments\,\cite{Konig_2007,Hsieh_2008,Hsieh_2009,Xia_2009,Chen_2009,Hsieh_2009}. Although most work considers band semiconductors, the concept of topology can be extended to correlated systems for which many exciting new effects can be expected. In analogy to correlation driven fractional quantum Hall states one might expect fractional Chern insulating states to emerge in correlated topological 
insulators\,\cite{Bergholtz_2013,Liu_2013,Flach_2014,Kourtis_2014,Grushin_2014}. The experimental realization of such states would open a whole new field of possibilities.

SmB$_6$ was recently reported to be a correlated mixed valent topological 
insulator\,\cite{Dzero_2010,Dzero_2012,Takimoto_2011,Wolgast_2013,Kim_2014,Suga_2014}. The highly correlated flat Sm-$f$ derived bands hybridize with the dispersive and itinerant Sm-$d$ derived bands to form a mixed valent insulating ground state. Although the insulating state and topology of the bulk of SmB$_6$ is well defined, [001] surface of
SmB$_{6}$ is polar and therefore inherently unstable\,\cite{Zhu_2013}. This hinders the unique determination of topological surface states and in turn inhibits the creation of robust technological devices. Here, using density functional theory (DFT) based calculations, we show that SmO is a mixed valent correlated compound with a band structure similar to the topological nontrivial high-pressure gold phase of SmS\,\cite{Li_2014}. In contrast to SmS, SmO is predicted to have a topological nontrivial ground state at ambient pressure. Additionally, our calculations show that the topological nontrivial ground-state of SmO is stable for a large range of Sm-O distances, including both positive and negative strain. The Sm-$f$ band filling is thus tunable by strain, which opens up the possibility to create correlated topological nontrivial bands at different filling. Subsequent to the enormous success in the design and fabrication of semiconductors, it is presently possible to grow high-quality oxide thin films and 
heterostructures\,\cite{Herber_2009,Mannhart_2010}. SmO thus seems the ideal candidate for the experimental realization of topological nontrivial correlated flat bands.
In addition, experimental realization of the theoretically predicted quantum anomalous Hall effect (QAHE)
(quantized Hall conductance in the absence of an external magnetic field) has been challenging
due to difficulties in obtaining insulating bulk TI's concurrent with homogeneous magnetic doping\,\cite{Ji_2012}.
Recently, notwithstanding the presence of bulk carrier density ($i.e.$ not a nominally insulating bulk), long range ferromagnetism and QAHE was successfully observed for the first time in Cr-doped (Bi,Sb)$_{2}$Te$_{3}$ TI thin films\,\cite{Chang_2013,Ke_2013}.
Other proposals towards realizing a QAHE are to either grow a TI on top of a ferromagnetic 
insulator\,\cite{Fu_2009,Akhmerov_2009}; transition-metal oxide heterostructures,\cite{Cook_2014,Cai_2013}  or
to deposit a layer of heavy atoms with large spin-orbit coupling on a magnetic insulator\,\cite{Garrity_2013}.
Fortuitously, a well studied substrate with the same symmetry as SmO is readily available in the form of the 
ferromagnetic semiconductor 
EuO\,\cite{Matthias_1961,Nolting_1979,Adachi_1998,Steeneken_2002,Schmehl_2007,Sutarto_2009,Melville_2013}.
Consequently, we propose growing SmO/EuO thin film interfaces to test the feasibility of obtaining a QAHE
in a strongly correlated electron system. 

\begin{figure*}[t]
\begin{center}
  \includegraphics[clip,width=0.9\textwidth]{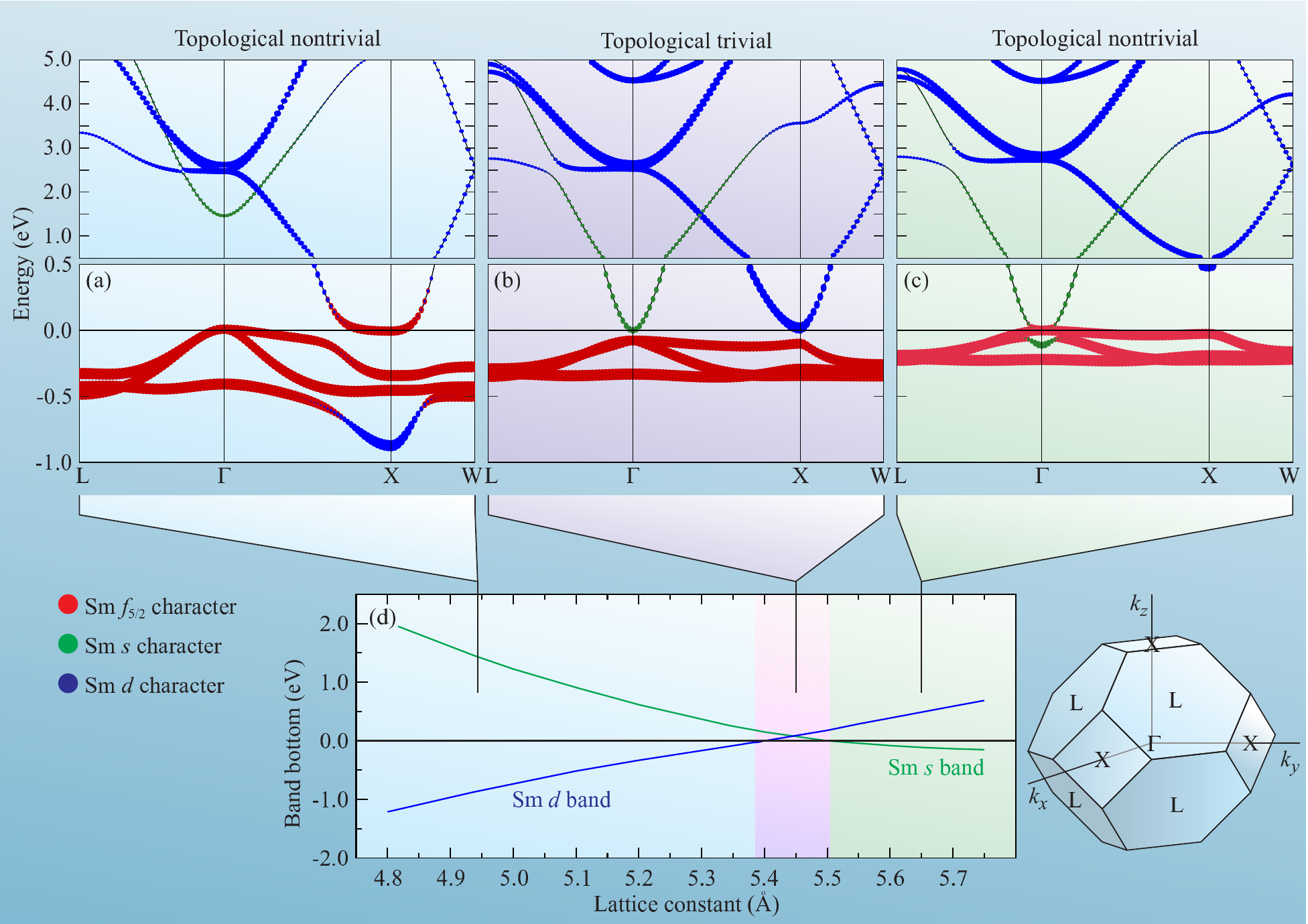}
\end{center}
\caption{(Color online) Phase diagram (d) of SmO as a function of the $fcc$ lattice parameter within LDA+SO+$U$ ($U$ = 6 eV, $J_{\mathrm{H}}$ = 0 eV) along with representative {\texttt{FPLO}} band structures for the (a) nontrivial
$d-f$ band inversion, (b) trivial insulator and (a) nontrivial $s-f$ band inversion scenarios.
The 4$f$, 5$d$ and 6$s$ orbital character derived bands are represented by red, blue and green colored 
symbols respectively. 
The size of the symbols represent the weight of the various orbital contributions to the underlying
bands. The Brillouin zone of an $fcc$ lattice is displayed along with the eight 
time reversal invariant momenta (TRIM) points.  }
\label{bands}
\end{figure*}

SmO crystallizes in the rock-salt structure
 with $a$\,=\,4.9414--4.943\,\AA\,\cite{Leger_1980,Krill_1980}.
Considering the lattice parameters of trivalent neodymium
chalcogenides and interpolating between neodymium and terbium
compounds, Leger {\it et. al.} obtain a lattice constant of
4.917\,\AA\, for Sm$^{3+}$O\,\cite{Leger_1980}. Similarly, considering
other divalent lanthanide oxides, a lattice constant of 5.15\,\AA\, is
expected for Sm$^{2+}$O.  Ergo, the experimental cell parameter of
$fcc$ SmO lies between that of Sm$^{3+}$O and Sm$^{2+}$O, which lead
the authors\,\cite{Leger_1980} to conclude that samarium is in an intermediate valence
state in SmO.  Using Vegard's law, a valence of 2.92 was
assigned\,\cite{Leger_1980}.
Electrical resistivity measurements as a function of temperature
reveals a $T^{2}$ like increase between 4.2 and 20\,K, followed by a
rapid increase up to 32 K. The resistivity shows a linear temperature
dependence above 60 K\,\cite{Leger_1980_a}.
Krill and co-workers\,\cite{Krill_1980}
measured the magnetic susceptibility of SmO and observed the roughly
constant magnetic susceptibility and the non-divergence of the
low-temperature susceptibility. They drew parallels to the
susceptibilities of other intermediate valence compounds SmB$_{6}$ and
gold-SmS under pressure, wherein the low temperature behavior is not
explainable by crystal field effects alone\,\cite{Krill_1980}.

The electronic structure calculations are performed using the
full-potential non-orthogonal local orbital code
(\texttt{FPLO})\,\cite{Koepernik_1999}. The local density approximation
(LDA) with the Perdew and Wang
flavor\,\cite{Perdew_1992} of the exchange and correlation potential was chosen. To
account for the strong spin-orbit (SO) coupling of the $f$ electrons, we
employ the full four-component relativistic scheme. Additionally, the
strong Coulomb repulsion between the 4$f$ electrons of samarium are included in a
mean-field way by applying LDA+SO+$U$ with the "fully localized limit" 
(FLL) double
counting term\,\cite{Czyzyk_1994}.  
This level of theory will not suffice to describe quantitatively the Sm 4$f$ spectral weight and its dispersion, but is sufficient to predict the Sm $d-f$ inversion. We have tested the robustness of these predictions by varying $U$ (5 to 7 eV) and $J_{\mathrm{H}}$ (0 to 0.7 eV) and by using different functionals (LDA and GGA using \texttt{FPLO} and the modified Becke-Johnson approach\cite{Tran_2009} with $U$=3 eV and $J_{\mathrm{H}}$=0 eV using \texttt{Wien2k}\cite{Blaha_2003}). Importantly, as pointed out by Martin and Allen\cite{Martin_1979} for SmB$_{6}$ and
related systems, the symmetry of the relevant Sm one particle orbitals and the many body electron removal Green's function is the same. This allows for a possible adiabatic continuation from the LDA results as presented here to the full interacting system without changing the topology of the system.  See Supplemental Material\,\footnote{See Supplemental Material at http://link.aps.org/Supplemental/...} for further discussion on the Sm 
4$f$ many body state.

\begin{figure}[t]
\begin{center}
  \includegraphics[clip,width=0.527\columnwidth]{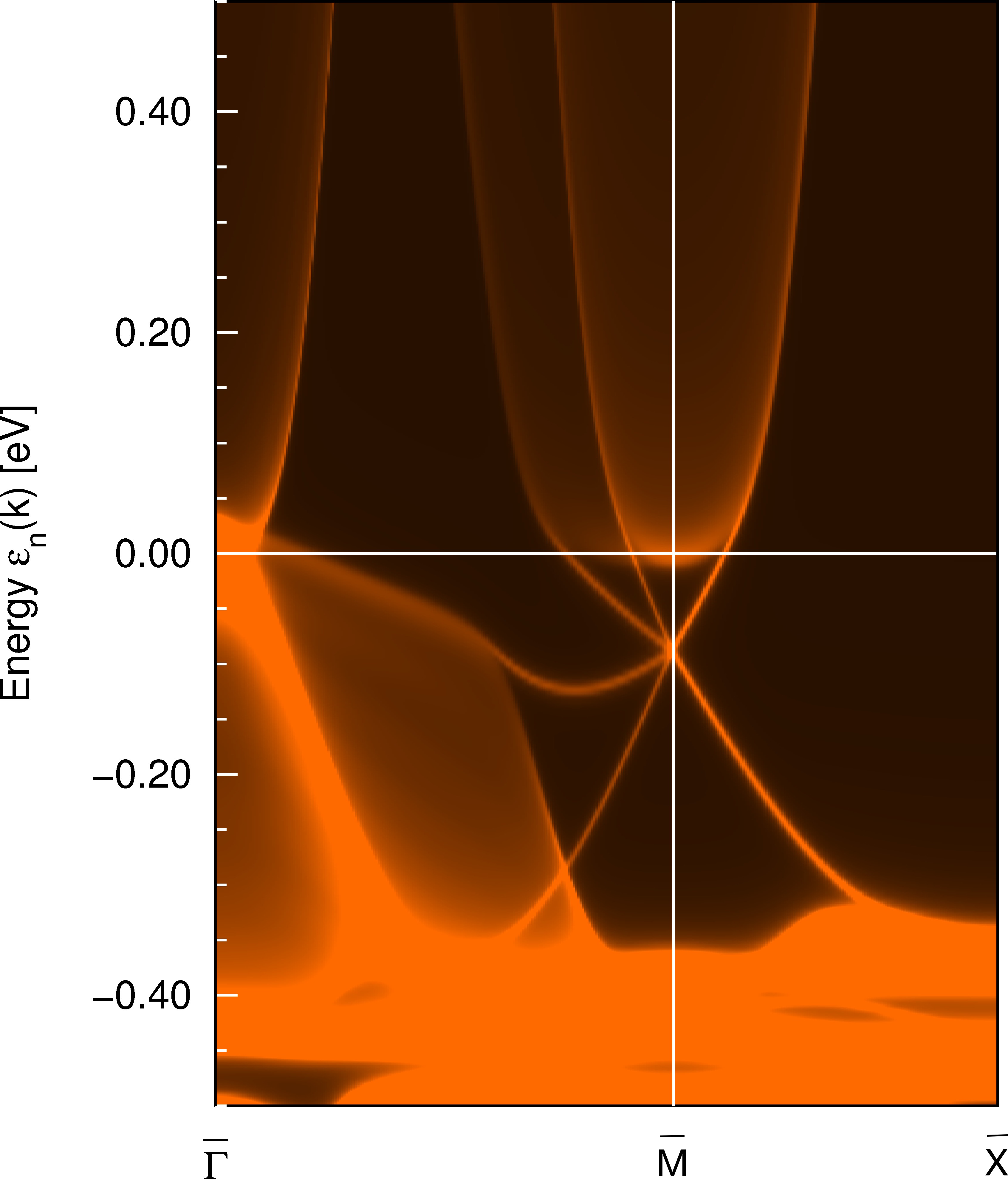}
   \includegraphics[clip,width=0.423\columnwidth]{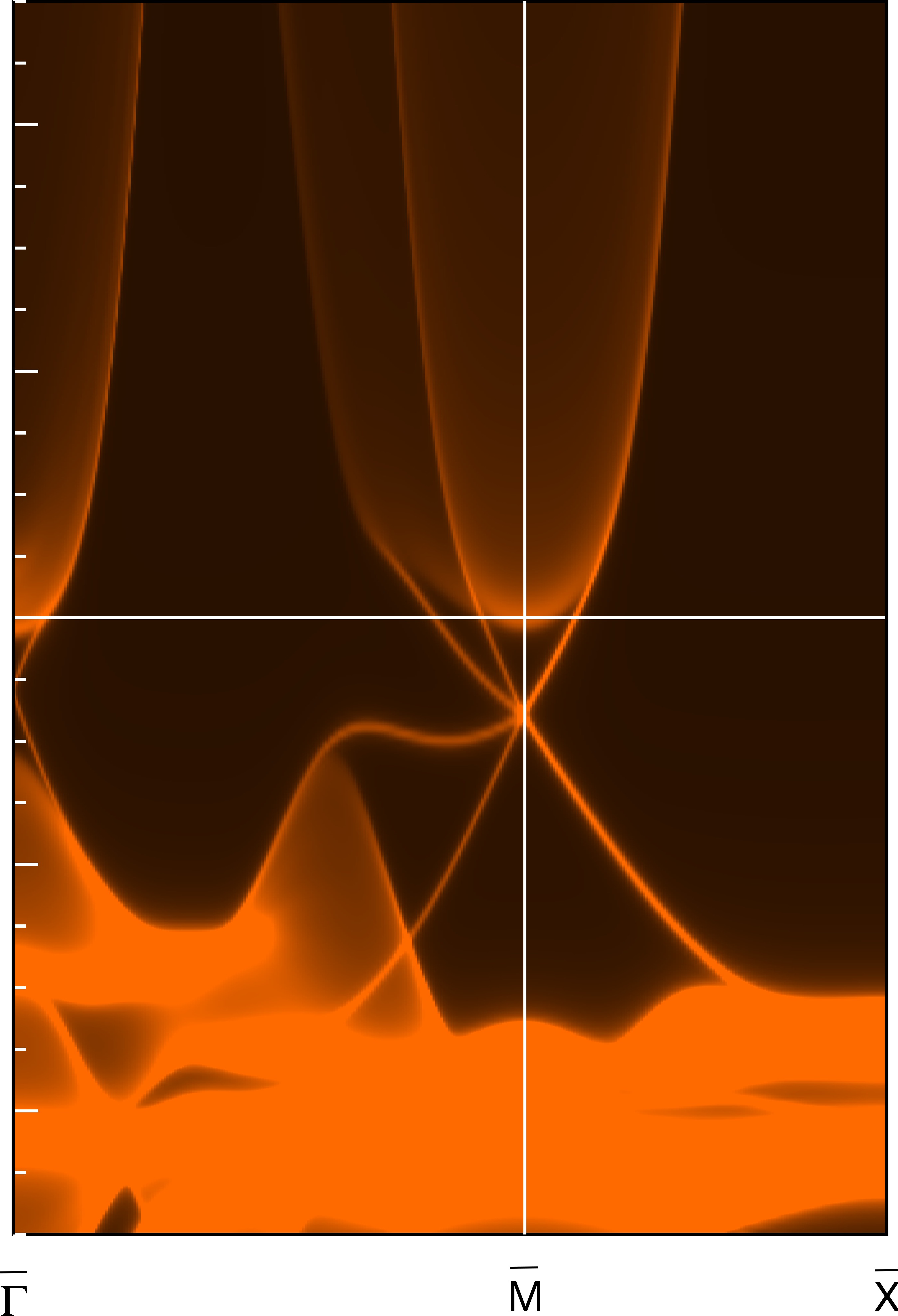}
\end{center}
\caption{(Color online) Left: Surface Bloch spectral density ($A_{\mathrm{Bl}}(k)$) of the first 12 SmO layers of a semi-infinite solid with [001]-surface. 
Note, that there are two Dirac cones at the $\bar{X}$-point, while the one at the $\bar{\Gamma}$-point falls into the bulk projected band
structure and hence forms a surface resonance (see right panel).
Right: $A_{\mathrm{Bl}}(k)$ with downwards shifted $4f$-electron pocket around the $\bar{\Gamma}$-point to reveal the third Dirac cone.
    }
\label{slab}
\end{figure}

Collected in Fig.\,\ref{bands}(a) is the {\texttt{FPLO}} non-spin polarized, full-relativistic band
structure of SmO with the inclusion
of the strong Coulomb interaction (LDA+SO+$U$) for the experimental lattice constant.  
With samarium being in the 2+ configuration, the
4$f$ states are split into lower lying and filled 4$f_{\mathrm{5/2}}$
states that can accommodate 6 electrons and higher lying (above 5.5 eV), empty 4$f_{\mathrm{7/2}}$ 
states (see Supplemental Material).
We do not observe a direct
energy gap around the Fermi level (E$_{\mathrm{F}}$). The material is semi-metallic with
a ``warped gap'' such that everywhere in the BZ (at E$_{\mathrm{F}}$),
band number $x$ is always below band number $x$+1.
There are 
small electron and hole pockets at X and $\Gamma$ respectively, but there are no band crossings
between the highest occupied 4$f_{\mathrm{5/2}}$ bands and the lowest
unoccupied bands.
Besides, varying $U$ (5 to 7 eV) or $J_{\mathrm{H}}$ (0 to 0.7 eV) does not change the above mentioned features, since
the oxygen 2$p$, samarium 5$d$ and 6$s$ bands experience a constant shift with respect 
to the localized 4$f$ states.   
 There exists only one
report\,\cite{Leger_1980_a} in literature on the transport
properties of SmO wherein the resistivity decreases with decreasing
temperature and thus could hint to metallic character, consistent with
the lack of an energy gap in our calculations.  
 In contrast to the localized and not very dispersive 4$f$ bands, the 5$d$
bands of samarium are dispersive, broad and dip below the Fermi level,
retaining a 100\% weight at the X point.  As a consequence, we observe
a 4$f$-5$d$ band inversion at the X point, resulting in a nontrivial
topology, similar in sense to the band inversions reported for 
SmB$_{6}$ and the high pressure gold phase of SmS. 
This band inversion and the resulting topological indices 1;(000) 
allows us
to classify SmO as a 3D strongly topological semi-metal
(more details are provided in Supplemental Material).

A well known issue with LDA when dealing with semiconductors, is the 
underestimation of band gaps. Since nontrivial topology depends on 
band inversions, underestimation of band gaps could sometimes lead to wrong sequence
of the underlying band structure. For example,  in HgTe, a $s$-$p$ TI, 
LDA correctly predicts the band inversion, but results in an incorrect band sequence at the $\Gamma$ point.
The correct band sequence is obtained by using the MBJLDA exchange potential\cite{Zhang_2013}.
Literature on the efficiency of MBJLDA over the traditional LDA+$U$ for $f$ electron systems is
still scarce. In a recent work on SmS, the authors have employed MBJLDA+SO+$U$ ($U$ = 3 eV)
to open a band gap of 0.2 eV in the ambient pressure black phase such that the
gap is consistent with the available experiments\,\cite{Li_2014}. Using the same parameters, the
authors conclude that the high pressure gold phase of SmS is a topological metal. 
We compared the LDA+SO+$U$ results with 
that of MBJLDA+SO and obtain a consistent picture  for the
band inversion between the 
two approaches (see Supplemental Material).
No direct band gap is opened with MBJLDA+SO+$U$ ($U$ = 3 eV), in accordance with our
LDA+SO+$U$ results and as well as with the experimental report\,\cite{Leger_1980_a}.

\begin{figure}[t]
 \includegraphics[width=\columnwidth]{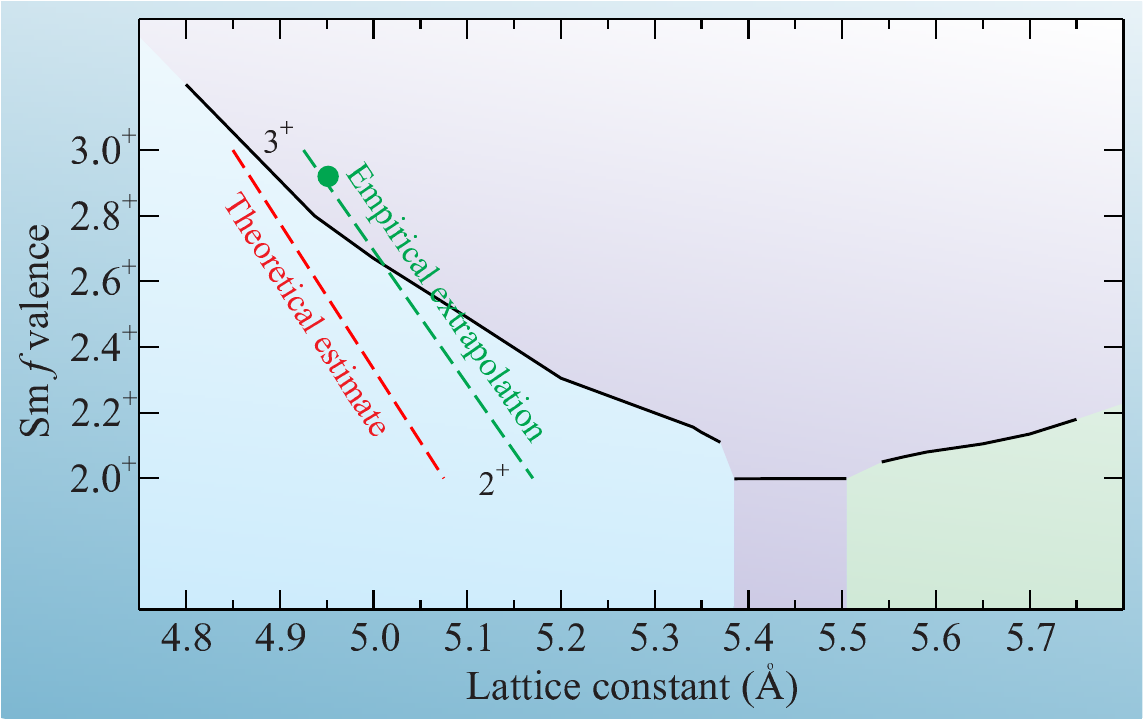}
\caption{(Color online) Maximum possible $f$ valence due to $f \longrightarrow d$ or
$f \longrightarrow s$ promotion in SmO while retaining the 
nontrivial topology as a function of
lattice parameter. 
The color map follows that of Fig.\,\ref{bands}(d): light blue = nontrivial topology from
$d-f$ inversion, light green = nontrivial topology from $s-f$ inversion, light pink = trivial
topology.
The equilibrium lattice parameters for a
high-spin 3+ and 2+ state are computed using L(S)DA+SO+$U$ scheme (dashed red line).
The estimates from an empirical extrapolation for the limiting cases are obtained from Ref.\,\onlinecite{Leger_1980} (dashed
green line). The filled green circle denotes the experimental bulk SmO.
 SmO remains a topological semi-metal for a wide range of 
tensile strain ($\geq$ 1\%), since the calculated maximum valence,
is above both the empirical extrapolation and L(S)DA+U+SO estimates. }\label{elects}
\end{figure}

To provide additional confirmation of the topological non-triviality in SmO and to explicitly
identify the protected surface states, we have calculated the Bloch spectral density 
($A_{\mathrm{Bl}}(k)$) of the 12 topmost surface layers of a semi infinite solid\cite{Velev_2007} with [001] surface.
We obtain two weakly interacting quasi-degenerate Dirac-cones at $\bar{M}$, and a single Dirac-cone at $\bar{\Gamma}$ which is hidden in the bulk
and becomes a surface resonance due to the semi-metallic nature of SmO. 
To clearly identify the Dirac-cone at $\bar{\Gamma}$, we have repeated the slab
calculations such that the hole pockets at $\Gamma$ for the bulk are pushed down.
This in turn, allows one to unambiguously identify all three Dirac-cones 
in a [001] terminated SmO. More details are provided in the Supplemental Material.

Having established the nontrivial topology in SmO for the experimental bulk
lattice parameter, we consider the scenario of growing thin films of SmO. 
In general, the lattice parameter of the substrate plays a decisive 
role in determining the lattice parameter of the thin film. 
Then, the relevant question to answer is the robustness of the topological semi-metal
state as a function of lattice parameter variation. 
To this end, we have investigated the topological indices for various lattice
parameters. We have retained the cubic symmetry of the unit cell ($i.e$ isotropic volume 
change) in SmO, based on the experimental studies on isostructural samarium systems, which
evidences a negative value of the Poisson ratio $\nu$ (elastic constant $C_{\mathrm{12}}$ < 0), 
characteristic of an isotropic volume change\cite{Luthi_1985,Wachter_1994}.
 Since the nontrivial topology is obtained due to $d-f$ band inversion at the X point, SmO is
topological as long as the 5$d$ band bottom is below the E$_{\mathrm{F}}$. The nontrivial topology is
suppressed when the 5$d$ band bottom moves above E$_{\mathrm{F}}$ or when other
bands dip below E$_{\mathrm{F}}$ (Fig.\,\ref{bands}(b)). In the case of SmO, it is the 6$s$ band bottom at the $\Gamma$ 
point that shifts to lower values as a function of increasing lattice parameter.
The shift of the 5$d$ and 6$s$ band bottoms (at X and $\Gamma$ respectively) with respect to the E$_{\mathrm{F}}$ as a function 
of lattice parameter are collected in Fig.\,\ref{bands}(d) for LDA+SO+$U$ approach.
Treating the strong 4$f$ correlations  on 
a mean-field level,  the topological semi-metal state is quite robust and remains so, up to 5.38\,\AA\,,
a 9\% increase in lattice parameter. On an experimental level, this result is very promising,
since it provides a large range of substrate lattice constants for which SmO thin films
reveal topological non-triviality.
During the evaluation of the topological indices using LDA+SO+$U$, we became aware of 
another interesting feature, a nontrivial topology due to $s-f$ band inversion at $\Gamma$ for 11\%
and larger lattice constants ($\geq$ 5.5\,\AA).  In LDA+SO+$U$, the 5$d$ band bottom shifts 
above the E$_{\mathrm{F}}$ and
becomes trivial before the 6$s$ band bottom dips below E$_{\mathrm{F}}$. 
Further increase in the lattice parameter then results in a $s-f$ band inversion at $\Gamma$
which 
is again topological (Fig.\,\ref{bands}(c)). 
Although, the probability for a successful growth of SmO on a substrate with a 
11\% and larger lattice constant may be low, it nonetheless offers another exciting route towards 
realizing a topologically nontrivial state, only this time with a $s-f$ band inversion. 
We verified the validity of our results for a range of $U$ values ( 5 to 7 eV) and as well as for
the MBJLDA exchange potential.

In the calculations so far, we have allowed Sm to be in the
intermediate valence state and have treated this on a DFT level. 
The mixed valent situation can be viewed as 
a 4$f$ to 5$d$ promotion of a certain fractional amount of electrons due to the hybridization 
of the localized 4$f$ states with the band like 5$d$ states.
Based on the handful of experimental reports available in literature, the valence assigned to Sm in SmO is 2.92,
$i.e.$ a promotion of 0.92 electrons. 
To actually calculate the amount of electron promotion is a difficult task,
since DFT in the Kohn-Sham scheme is based on a single Slater determinant approximation and prevents a 
 correct description of a many body intermediate valence state.  
In other words, the position of the 4$f$ bands relative to the 5$d$/6$s$ as calculated in DFT is
not a reliable quantity (see Supplemental Material). Yet, we can address the effect of mixed valency on the topological
properties of SmO
 in a rigorous quantitative way, by calculating the amount of electrons (integration
of the density of states) that
can be accommodated in the dispersive 5$d$ or 6$s$ band, thereby
mimicking various amounts of 4$f$ $\longrightarrow$ 5$d$ or 4$f$ $\longrightarrow$ 6$s$ promotion.
For the lattice constants which has the band inversion at the X point,
 we calculate the amount of electrons that can be contained in the 5$d$ band before beginning
to populate the 6$s$ band. Note that the nontrivial topology is maintained as long as only
the 5$d$ is occupied $and$ the 6$s$ remains unoccupied. Fig.\,\ref{elects} displays this electron amount as
a function of the lattice constant.
Analogously, on the other side of the phase diagram, wherein non-triviality is manifested due to a $s-f$ inversion, 
we calculate the amount of electrons that can be contained in the 6$s$ band before beginning to 
populate the 5$d$ band.
Equilibrium lattice constants for the 3+ and 2+ limiting cases are estimated using spin-polarized L(S)DA+SO+$U$ schemes (see Supplemental Material).
We observe that the theoretical estimated valence and lattice constants always yield a topological
nontrivial ground state.
Though the estimate from an empirical extrapolation for the SmO valence places the bulk in the trivial
region of the phase diagram, nontrivial topology is quickly reinstated for lattice parameters
with a small tensile strain  ($\geq$ 1\%). 
Spectroscopy experiments are necessary to confirm the samarium 
valency, and positions of the samarium $s$ and $d$ bands with respect to the $f$ states in bulk SmO and SmO thin films.

By virtue of the large surface to volume ratio, thin films of systems with nontrivial topology with 
dominant surface states are highly sought-after.  
Owing to the simple $fcc$ symmetry and the flexibility in observing the topological ground state in SmO for a wide
range of lattice parameters, we anticipate plenty of options for suitable substrates. 
One particular substrate that invokes special interest is the $fcc$ ferromagnetic semiconductor 
EuO\,\cite{Matthias_1961,Nolting_1979,Adachi_1998,Steeneken_2002,Schmehl_2007,Sutarto_2009,Melville_2013},
which has a lattice constant of 5.14\,\AA, only a 4\% lattice mismatch with that of bulk SmO.  
EuO by itself is an attractive material with many functionalities, including metal-insulator transition, magnetic
phase transition, colossal magneto-resistance, 
etc\,\cite{Steeneken_2002,Schmehl_2007,Sutarto_2009,Melville_2013}.
These functionalities have produced an abundance of research on growing high quality thin films of EuO.
So, a quick progress in the engineering of SmO/EuO thin film interfaces can be expected. 
Additionally, this would open up the possibility for another experimental realization of the QAHE,
this time in a strongly correlated electron system with topologically nontrivial flat bands. 
Despite the lack of a bulk band gap in Cr-doped (Bi,Sb)$_{2}$Te$_{3}$ thin films,
QAHE was observed at 30 mK\,\cite{Chang_2013,Ke_2013}. In a similar manner, the presence of a semi-metallic bulk in SmO will give rise to bulk carriers, but given the tiny electron and hole
pockets around X and $\Gamma$ respectively, and by making use of thin films, 
the amount of bulk charge carriers can be kept
quite small and should not completely overshadow the QAHE.

In summary, we have investigated in detail the electronic structure of SmO and established
the nontrivial topology of the band structure such that SmO can be 
classified as a 3D strongly topological semi-metal. 
The topological state in SmO is very robust and prevails for a wide range of lattice parameters,
which leads us to propose SmO as an ideal candidate for further investigations of topologically nontrivial
correlated flat bands in bulk and thin-film form.

\begin{acknowledgments}
The authors thank Prof. J. Allen, S. Wirth and P. Thalmeier for discussions. D.K. acknowledges
funding by the DFG within FOR 1346. 
\end{acknowledgments}


\begin{thebibliography}{56}%
\makeatletter
\providecommand \@ifxundefined [1]{%
 \@ifx{#1\undefined}
}%
\providecommand \@ifnum [1]{%
 \ifnum #1\expandafter \@firstoftwo
 \else \expandafter \@secondoftwo
 \fi
}%
\providecommand \@ifx [1]{%
 \ifx #1\expandafter \@firstoftwo
 \else \expandafter \@secondoftwo
 \fi
}%
\providecommand \natexlab [1]{#1}%
\providecommand \enquote  [1]{``#1''}%
\providecommand \bibnamefont  [1]{#1}%
\providecommand \bibfnamefont [1]{#1}%
\providecommand \citenamefont [1]{#1}%
\providecommand \href@noop [0]{\@secondoftwo}%
\providecommand \href [0]{\begingroup \@sanitize@url \@href}%
\providecommand \@href[1]{\@@startlink{#1}\@@href}%
\providecommand \@@href[1]{\endgroup#1\@@endlink}%
\providecommand \@sanitize@url [0]{\catcode `\\12\catcode `\$12\catcode
  `\&12\catcode `\#12\catcode `\^12\catcode `\_12\catcode `\%12\relax}%
\providecommand \@@startlink[1]{}%
\providecommand \@@endlink[0]{}%
\providecommand \url  [0]{\begingroup\@sanitize@url \@url }%
\providecommand \@url [1]{\endgroup\@href {#1}{\urlprefix }}%
\providecommand \urlprefix  [0]{URL }%
\providecommand \Eprint [0]{\href }%
\providecommand \doibase [0]{http://dx.doi.org/}%
\providecommand \selectlanguage [0]{\@gobble}%
\providecommand \bibinfo  [0]{\@secondoftwo}%
\providecommand \bibfield  [0]{\@secondoftwo}%
\providecommand \translation [1]{[#1]}%
\providecommand \BibitemOpen [0]{}%
\providecommand \bibitemStop [0]{}%
\providecommand \bibitemNoStop [0]{.\EOS\space}%
\providecommand \EOS [0]{\spacefactor3000\relax}%
\providecommand \BibitemShut  [1]{\csname bibitem#1\endcsname}%
\let\auto@bib@innerbib\@empty
\bibitem [{\citenamefont {Kane}\ and\ \citenamefont
  {Mele}(2005{\natexlab{a}})}]{Kane_2005}%
  \BibitemOpen
  \bibfield  {author} {\bibinfo {author} {\bibfnamefont {C.~L.}\ \bibnamefont
  {Kane}}\ and\ \bibinfo {author} {\bibfnamefont {E.~J.}\ \bibnamefont
  {Mele}},\ }\href@noop {} {\bibfield  {journal} {\bibinfo  {journal} {Phys.
  Rev. Lett.}\ }\textbf {\bibinfo {volume} {95}},\ \bibinfo {pages} {146802}
  (\bibinfo {year} {2005}{\natexlab{a}})}\BibitemShut {NoStop}%
\bibitem [{\citenamefont {Kane}\ and\ \citenamefont
  {Mele}(2005{\natexlab{b}})}]{Kane_2005_2}%
  \BibitemOpen
  \bibfield  {author} {\bibinfo {author} {\bibfnamefont {C.~L.}\ \bibnamefont
  {Kane}}\ and\ \bibinfo {author} {\bibfnamefont {E.~J.}\ \bibnamefont
  {Mele}},\ }\href@noop {} {\bibfield  {journal} {\bibinfo  {journal} {Phys.
  Rev. Lett.}\ }\textbf {\bibinfo {volume} {95}},\ \bibinfo {pages} {226801}
  (\bibinfo {year} {2005}{\natexlab{b}})}\BibitemShut {NoStop}%
\bibitem [{\citenamefont {Fu}\ and\ \citenamefont {Kane}(2007)}]{Fu_2007}%
  \BibitemOpen
  \bibfield  {author} {\bibinfo {author} {\bibfnamefont {L.}~\bibnamefont
  {Fu}}\ and\ \bibinfo {author} {\bibfnamefont {C.~L.}\ \bibnamefont {Kane}},\
  }\href@noop {} {\bibfield  {journal} {\bibinfo  {journal} {Phys. Rev. B}\
  }\textbf {\bibinfo {volume} {76}},\ \bibinfo {pages} {045302} (\bibinfo
  {year} {2007})}\BibitemShut {NoStop}%
\bibitem [{\citenamefont {Moore}\ and\ \citenamefont
  {Balents}(2007)}]{Moore_2007}%
  \BibitemOpen
  \bibfield  {author} {\bibinfo {author} {\bibfnamefont {J.~E.}\ \bibnamefont
  {Moore}}\ and\ \bibinfo {author} {\bibfnamefont {L.}~\bibnamefont
  {Balents}},\ }\href@noop {} {\bibfield  {journal} {\bibinfo  {journal} {Phys.
  Rev. B}\ }\textbf {\bibinfo {volume} {75}},\ \bibinfo {pages} {121306(R)}
  (\bibinfo {year} {2007})}\BibitemShut {NoStop}%
\bibitem [{\citenamefont {Roy}(2009)}]{Roy_2009}%
  \BibitemOpen
  \bibfield  {author} {\bibinfo {author} {\bibfnamefont {R.}~\bibnamefont
  {Roy}},\ }\href@noop {} {\bibfield  {journal} {\bibinfo  {journal} {Phys.
  Rev. B}\ }\textbf {\bibinfo {volume} {79}},\ \bibinfo {pages} {195321}
  (\bibinfo {year} {2009})}\BibitemShut {NoStop}%
\bibitem [{\citenamefont {Fu}\ \emph {et~al.}(2007)\citenamefont {Fu},
  \citenamefont {Kane},\ and\ \citenamefont {Mele}}]{Fu_2007_2}%
  \BibitemOpen
  \bibfield  {author} {\bibinfo {author} {\bibfnamefont {L.}~\bibnamefont
  {Fu}}, \bibinfo {author} {\bibfnamefont {C.~L.}\ \bibnamefont {Kane}}, \ and\
  \bibinfo {author} {\bibfnamefont {E.~J.}\ \bibnamefont {Mele}},\ }\href@noop
  {} {\bibfield  {journal} {\bibinfo  {journal} {Phys. Rev. Lett.}\ }\textbf
  {\bibinfo {volume} {98}},\ \bibinfo {pages} {106803} (\bibinfo {year}
  {2007})}\BibitemShut {NoStop}%
\bibitem [{\citenamefont {Zhang}\ \emph {et~al.}(2009)\citenamefont {Zhang},
  \citenamefont {Liu}, \citenamefont {Qi}, \citenamefont {Dai}, \citenamefont
  {Fang},\ and\ \citenamefont {Zhang}}]{Zhang_2009}%
  \BibitemOpen
  \bibfield  {author} {\bibinfo {author} {\bibfnamefont {H.}~\bibnamefont
  {Zhang}}, \bibinfo {author} {\bibfnamefont {C.-X.}\ \bibnamefont {Liu}},
  \bibinfo {author} {\bibfnamefont {X.-L.}\ \bibnamefont {Qi}}, \bibinfo
  {author} {\bibfnamefont {X.}~\bibnamefont {Dai}}, \bibinfo {author}
  {\bibfnamefont {Z.}~\bibnamefont {Fang}}, \ and\ \bibinfo {author}
  {\bibfnamefont {S.-C.}\ \bibnamefont {Zhang}},\ }\href@noop {} {\bibfield
  {journal} {\bibinfo  {journal} {Nat. Phys.}\ }\textbf {\bibinfo {volume}
  {5}},\ \bibinfo {pages} {438} (\bibinfo {year} {2009})}\BibitemShut {NoStop}%
\bibitem [{\citenamefont {Konig}\ \emph {et~al.}(2007)\citenamefont {Konig},
  \citenamefont {Wiedmann}, \citenamefont {Brune}, \citenamefont {Roth},
  \citenamefont {Buhmann}, \citenamefont {Molenkamp}, \citenamefont {Qi},\ and\
  \citenamefont {Zhang}}]{Konig_2007}%
  \BibitemOpen
  \bibfield  {author} {\bibinfo {author} {\bibfnamefont {M.}~\bibnamefont
  {Konig}}, \bibinfo {author} {\bibfnamefont {S.}~\bibnamefont {Wiedmann}},
  \bibinfo {author} {\bibfnamefont {C.}~\bibnamefont {Brune}}, \bibinfo
  {author} {\bibfnamefont {A.}~\bibnamefont {Roth}}, \bibinfo {author}
  {\bibfnamefont {H.}~\bibnamefont {Buhmann}}, \bibinfo {author} {\bibfnamefont
  {L.~W.}\ \bibnamefont {Molenkamp}}, \bibinfo {author} {\bibfnamefont {X.~L.}\
  \bibnamefont {Qi}}, \ and\ \bibinfo {author} {\bibfnamefont {S.~C.}\
  \bibnamefont {Zhang}},\ }\href@noop {} {\bibfield  {journal} {\bibinfo
  {journal} {Science}\ }\textbf {\bibinfo {volume} {318}},\ \bibinfo {pages}
  {766} (\bibinfo {year} {2007})}\BibitemShut {NoStop}%
\bibitem [{\citenamefont {Hsieh}\ \emph {et~al.}(2008)\citenamefont {Hsieh},
  \citenamefont {Qian}, \citenamefont {Wray}, \citenamefont {Xia},
  \citenamefont {Hor}, \citenamefont {Cava},\ and\ \citenamefont
  {Hasan}}]{Hsieh_2008}%
  \BibitemOpen
  \bibfield  {author} {\bibinfo {author} {\bibfnamefont {D.}~\bibnamefont
  {Hsieh}}, \bibinfo {author} {\bibfnamefont {D.}~\bibnamefont {Qian}},
  \bibinfo {author} {\bibfnamefont {L.}~\bibnamefont {Wray}}, \bibinfo {author}
  {\bibfnamefont {Y.}~\bibnamefont {Xia}}, \bibinfo {author} {\bibfnamefont
  {Y.~S.}\ \bibnamefont {Hor}}, \bibinfo {author} {\bibfnamefont {R.~J.}\
  \bibnamefont {Cava}}, \ and\ \bibinfo {author} {\bibfnamefont {M.~Z.}\
  \bibnamefont {Hasan}},\ }\href@noop {} {\bibfield  {journal} {\bibinfo
  {journal} {Nature}\ }\textbf {\bibinfo {volume} {452}},\ \bibinfo {pages}
  {970} (\bibinfo {year} {2008})}\BibitemShut {NoStop}%
\bibitem [{\citenamefont {Hsieh}\ \emph {et~al.}(2009)\citenamefont {Hsieh},
  \citenamefont {Xia}, \citenamefont {Wray}, \citenamefont {Qian},
  \citenamefont {Pal}, \citenamefont {Dil}, \citenamefont {Osterwalder},
  \citenamefont {Meier}, \citenamefont {Bihlmayer}, \citenamefont {Kane},
  \citenamefont {Hor}, \citenamefont {Cava},\ and\ \citenamefont
  {Z.}}]{Hsieh_2009}%
  \BibitemOpen
  \bibfield  {author} {\bibinfo {author} {\bibfnamefont {D.}~\bibnamefont
  {Hsieh}}, \bibinfo {author} {\bibfnamefont {Y.}~\bibnamefont {Xia}}, \bibinfo
  {author} {\bibfnamefont {L.}~\bibnamefont {Wray}}, \bibinfo {author}
  {\bibfnamefont {D.}~\bibnamefont {Qian}}, \bibinfo {author} {\bibfnamefont
  {A.}~\bibnamefont {Pal}}, \bibinfo {author} {\bibfnamefont {H.}~\bibnamefont
  {Dil}}, \bibinfo {author} {\bibfnamefont {J.}~\bibnamefont {Osterwalder}},
  \bibinfo {author} {\bibfnamefont {F.}~\bibnamefont {Meier}}, \bibinfo
  {author} {\bibfnamefont {G.}~\bibnamefont {Bihlmayer}}, \bibinfo {author}
  {\bibfnamefont {C.~L.}\ \bibnamefont {Kane}}, \bibinfo {author}
  {\bibfnamefont {Y.~S.}\ \bibnamefont {Hor}}, \bibinfo {author} {\bibfnamefont
  {R.~J.}\ \bibnamefont {Cava}}, \ and\ \bibinfo {author} {\bibfnamefont
  {H.~M.}\ \bibnamefont {Z.}},\ }\href@noop {} {\bibfield  {journal} {\bibinfo
  {journal} {Science}\ }\textbf {\bibinfo {volume} {323}},\ \bibinfo {pages}
  {919} (\bibinfo {year} {2009})}\BibitemShut {NoStop}%
\bibitem [{\citenamefont {Xia}\ \emph {et~al.}(2009)\citenamefont {Xia},
  \citenamefont {Qian}, \citenamefont {Hsieh}, \citenamefont {Wray},
  \citenamefont {Pal}, \citenamefont {Lin}, \citenamefont {Bansil},
  \citenamefont {Grauer}, \citenamefont {Hor}, \citenamefont {Cava},\ and\
  \citenamefont {Hasan}}]{Xia_2009}%
  \BibitemOpen
  \bibfield  {author} {\bibinfo {author} {\bibfnamefont {Y.}~\bibnamefont
  {Xia}}, \bibinfo {author} {\bibfnamefont {D.}~\bibnamefont {Qian}}, \bibinfo
  {author} {\bibfnamefont {D.}~\bibnamefont {Hsieh}}, \bibinfo {author}
  {\bibfnamefont {L.}~\bibnamefont {Wray}}, \bibinfo {author} {\bibfnamefont
  {A.}~\bibnamefont {Pal}}, \bibinfo {author} {\bibfnamefont {H.}~\bibnamefont
  {Lin}}, \bibinfo {author} {\bibfnamefont {A.}~\bibnamefont {Bansil}},
  \bibinfo {author} {\bibfnamefont {D.}~\bibnamefont {Grauer}}, \bibinfo
  {author} {\bibfnamefont {Y.~S.}\ \bibnamefont {Hor}}, \bibinfo {author}
  {\bibfnamefont {R.~J.}\ \bibnamefont {Cava}}, \ and\ \bibinfo {author}
  {\bibfnamefont {M.~Z.}\ \bibnamefont {Hasan}},\ }\href@noop {} {\bibfield
  {journal} {\bibinfo  {journal} {Nat. Phys.}\ }\textbf {\bibinfo {volume}
  {5}},\ \bibinfo {pages} {398} (\bibinfo {year} {2009})}\BibitemShut {NoStop}%
\bibitem [{\citenamefont {Chen}\ \emph {et~al.}(2009)\citenamefont {Chen},
  \citenamefont {Analytis}, \citenamefont {Chu}, \citenamefont {Liu},
  \citenamefont {Mo}, \citenamefont {Qi}, \citenamefont {Zhang}, \citenamefont
  {Lu}, \citenamefont {Dai}, \citenamefont {Fang}, \citenamefont {Zhang},
  \citenamefont {Fisher}, \citenamefont {Hussain},\ and\ \citenamefont
  {Shen}}]{Chen_2009}%
  \BibitemOpen
  \bibfield  {author} {\bibinfo {author} {\bibfnamefont {Y.~L.}\ \bibnamefont
  {Chen}}, \bibinfo {author} {\bibfnamefont {J.~G.}\ \bibnamefont {Analytis}},
  \bibinfo {author} {\bibfnamefont {J.-H.}\ \bibnamefont {Chu}}, \bibinfo
  {author} {\bibfnamefont {Z.~K.}\ \bibnamefont {Liu}}, \bibinfo {author}
  {\bibfnamefont {S.-K.}\ \bibnamefont {Mo}}, \bibinfo {author} {\bibfnamefont
  {X.~L.}\ \bibnamefont {Qi}}, \bibinfo {author} {\bibfnamefont {H.~J.}\
  \bibnamefont {Zhang}}, \bibinfo {author} {\bibfnamefont {D.~H.}\ \bibnamefont
  {Lu}}, \bibinfo {author} {\bibfnamefont {X.}~\bibnamefont {Dai}}, \bibinfo
  {author} {\bibfnamefont {Z.}~\bibnamefont {Fang}}, \bibinfo {author}
  {\bibfnamefont {S.~C.}\ \bibnamefont {Zhang}}, \bibinfo {author}
  {\bibfnamefont {I.~R.}\ \bibnamefont {Fisher}}, \bibinfo {author}
  {\bibfnamefont {Z.}~\bibnamefont {Hussain}}, \ and\ \bibinfo {author}
  {\bibfnamefont {Z.-X.}\ \bibnamefont {Shen}},\ }\href@noop {} {\bibfield
  {journal} {\bibinfo  {journal} {Science}\ }\textbf {\bibinfo {volume}
  {325}},\ \bibinfo {pages} {325} (\bibinfo {year} {2009})}\BibitemShut
  {NoStop}%
\bibitem [{\citenamefont {Bergholtz}\ and\ \citenamefont
  {Liu}(2013)}]{Bergholtz_2013}%
  \BibitemOpen
  \bibfield  {author} {\bibinfo {author} {\bibfnamefont {E.}~\bibnamefont
  {Bergholtz}}\ and\ \bibinfo {author} {\bibfnamefont {Z.}~\bibnamefont
  {Liu}},\ }\href@noop {} {\bibfield  {journal} {\bibinfo  {journal} {Int. J.
  Mod. Phys. B}\ }\textbf {\bibinfo {volume} {27}},\ \bibinfo {pages} {1330017}
  (\bibinfo {year} {2013})}\BibitemShut {NoStop}%
\bibitem [{\citenamefont {Liu}\ \emph {et~al.}(2013)\citenamefont {Liu},
  \citenamefont {Bergholtz},\ and\ \citenamefont {Kapit}}]{Liu_2013}%
  \BibitemOpen
  \bibfield  {author} {\bibinfo {author} {\bibfnamefont {Z.}~\bibnamefont
  {Liu}}, \bibinfo {author} {\bibfnamefont {E.~J.}\ \bibnamefont {Bergholtz}},
  \ and\ \bibinfo {author} {\bibfnamefont {E.}~\bibnamefont {Kapit}},\
  }\href@noop {} {\bibfield  {journal} {\bibinfo  {journal} {Phys. Rev. B}\
  }\textbf {\bibinfo {volume} {88}},\ \bibinfo {pages} {205101} (\bibinfo
  {year} {2013})}\BibitemShut {NoStop}%
\bibitem [{\citenamefont {Flach}\ \emph {et~al.}(2014)\citenamefont {Flach},
  \citenamefont {Leykam}, \citenamefont {Bodyfelt}, \citenamefont {Matthies},\
  and\ \citenamefont {Desyatnikov}}]{Flach_2014}%
  \BibitemOpen
  \bibfield  {author} {\bibinfo {author} {\bibfnamefont {S.}~\bibnamefont
  {Flach}}, \bibinfo {author} {\bibfnamefont {D.}~\bibnamefont {Leykam}},
  \bibinfo {author} {\bibfnamefont {J.~D.}\ \bibnamefont {Bodyfelt}}, \bibinfo
  {author} {\bibfnamefont {P.}~\bibnamefont {Matthies}}, \ and\ \bibinfo
  {author} {\bibfnamefont {A.~S.}\ \bibnamefont {Desyatnikov}},\ }\href@noop {}
  {\bibfield  {journal} {\bibinfo  {journal} {Euro Phys. Lett.}\ }\textbf
  {\bibinfo {volume} {105}},\ \bibinfo {pages} {30001} (\bibinfo {year}
  {2014})}\BibitemShut {NoStop}%
\bibitem [{\citenamefont {Kourtis}\ \emph {et~al.}(2014)\citenamefont
  {Kourtis}, \citenamefont {Neupert}, \citenamefont {Chamon},\ and\
  \citenamefont {Mudry}}]{Kourtis_2014}%
  \BibitemOpen
  \bibfield  {author} {\bibinfo {author} {\bibfnamefont {S.}~\bibnamefont
  {Kourtis}}, \bibinfo {author} {\bibfnamefont {T.}~\bibnamefont {Neupert}},
  \bibinfo {author} {\bibfnamefont {C.}~\bibnamefont {Chamon}}, \ and\ \bibinfo
  {author} {\bibfnamefont {C.}~\bibnamefont {Mudry}},\ }\href@noop {}
  {\bibfield  {journal} {\bibinfo  {journal} {Phys. Rev. Lett.}\ }\textbf
  {\bibinfo {volume} {112}},\ \bibinfo {pages} {126806} (\bibinfo {year}
  {2014})}\BibitemShut {NoStop}%
\bibitem [{\citenamefont {Grushin}\ \emph {et~al.}(2014)\citenamefont
  {Grushin}, \citenamefont {G{\'{o}}mez-Le{\'{o}}n},\ and\ \citenamefont
  {Neupert}}]{Grushin_2014}%
  \BibitemOpen
  \bibfield  {author} {\bibinfo {author} {\bibfnamefont {A.~G.}\ \bibnamefont
  {Grushin}}, \bibinfo {author} {\bibfnamefont {{\'{A}}.}~\bibnamefont
  {G{\'{o}}mez-Le{\'{o}}n}}, \ and\ \bibinfo {author} {\bibfnamefont
  {T.}~\bibnamefont {Neupert}},\ }\href@noop {} {\bibfield  {journal} {\bibinfo
   {journal} {Phys. Rev. Lett.}\ }\textbf {\bibinfo {volume} {112}},\ \bibinfo
  {pages} {156801} (\bibinfo {year} {2014})}\BibitemShut {NoStop}%
\bibitem [{\citenamefont {Dzero}\ \emph {et~al.}(2010)\citenamefont {Dzero},
  \citenamefont {Sun}, \citenamefont {Galitski},\ and\ \citenamefont
  {Coleman}}]{Dzero_2010}%
  \BibitemOpen
  \bibfield  {author} {\bibinfo {author} {\bibfnamefont {M.}~\bibnamefont
  {Dzero}}, \bibinfo {author} {\bibfnamefont {K.}~\bibnamefont {Sun}}, \bibinfo
  {author} {\bibfnamefont {V.}~\bibnamefont {Galitski}}, \ and\ \bibinfo
  {author} {\bibfnamefont {P.}~\bibnamefont {Coleman}},\ }\href@noop {}
  {\bibfield  {journal} {\bibinfo  {journal} {Phys. Rev. Lett.}\ }\textbf
  {\bibinfo {volume} {104}},\ \bibinfo {pages} {106408} (\bibinfo {year}
  {2010})}\BibitemShut {NoStop}%
\bibitem [{\citenamefont {Dzero}\ \emph {et~al.}(2012)\citenamefont {Dzero},
  \citenamefont {Sun}, \citenamefont {Coleman},\ and\ \citenamefont
  {V.}}]{Dzero_2012}%
  \BibitemOpen
  \bibfield  {author} {\bibinfo {author} {\bibfnamefont {M.}~\bibnamefont
  {Dzero}}, \bibinfo {author} {\bibfnamefont {K.}~\bibnamefont {Sun}}, \bibinfo
  {author} {\bibfnamefont {P.}~\bibnamefont {Coleman}}, \ and\ \bibinfo
  {author} {\bibfnamefont {G.}~\bibnamefont {V.}},\ }\href@noop {} {\bibfield
  {journal} {\bibinfo  {journal} {Phys. Rev. B}\ }\textbf {\bibinfo {volume}
  {85}},\ \bibinfo {pages} {045130} (\bibinfo {year} {2012})}\BibitemShut
  {NoStop}%
\bibitem [{\citenamefont {Takimoto}(2011)}]{Takimoto_2011}%
  \BibitemOpen
  \bibfield  {author} {\bibinfo {author} {\bibfnamefont {T.}~\bibnamefont
  {Takimoto}},\ }\href@noop {} {\bibfield  {journal} {\bibinfo  {journal} {J.
  Phys. Soc. Jpn.}\ }\textbf {\bibinfo {volume} {80}},\ \bibinfo {pages}
  {123710} (\bibinfo {year} {2011})}\BibitemShut {NoStop}%
\bibitem [{\citenamefont {Wolgast}\ \emph {et~al.}(2013)\citenamefont
  {Wolgast}, \citenamefont {Kurdak}, \citenamefont {Sun}, \citenamefont
  {Allen}, \citenamefont {Kim},\ and\ \citenamefont {Fisk}}]{Wolgast_2013}%
  \BibitemOpen
  \bibfield  {author} {\bibinfo {author} {\bibfnamefont {S.}~\bibnamefont
  {Wolgast}}, \bibinfo {author} {\bibfnamefont {t.}~\bibnamefont {Kurdak}},
  \bibinfo {author} {\bibfnamefont {K.}~\bibnamefont {Sun}}, \bibinfo {author}
  {\bibfnamefont {J.~W.}\ \bibnamefont {Allen}}, \bibinfo {author}
  {\bibfnamefont {D.-J.}\ \bibnamefont {Kim}}, \ and\ \bibinfo {author}
  {\bibfnamefont {Z.}~\bibnamefont {Fisk}},\ }\href@noop {} {\bibfield
  {journal} {\bibinfo  {journal} {Phys. Rev. B}\ }\textbf {\bibinfo {volume}
  {88}},\ \bibinfo {pages} {180405(R)} (\bibinfo {year} {2013})}\BibitemShut
  {NoStop}%
\bibitem [{\citenamefont {Kim}\ \emph {et~al.}(2014)\citenamefont {Kim},
  \citenamefont {Xia},\ and\ \citenamefont {Fisk}}]{Kim_2014}%
  \BibitemOpen
  \bibfield  {author} {\bibinfo {author} {\bibfnamefont {D.~J.}\ \bibnamefont
  {Kim}}, \bibinfo {author} {\bibfnamefont {J.}~\bibnamefont {Xia}}, \ and\
  \bibinfo {author} {\bibfnamefont {Z.}~\bibnamefont {Fisk}},\ }\href@noop {}
  {\bibfield  {journal} {\bibinfo  {journal} {Nature Mat.}\ }\textbf {\bibinfo
  {volume} {13}},\ \bibinfo {pages} {466} (\bibinfo {year} {2014})}\BibitemShut
  {NoStop}%
\bibitem [{\citenamefont {Suga}\ \emph {et~al.}(2014)\citenamefont {Suga},
  \citenamefont {Sakamoto}, \citenamefont {Okuda}, \citenamefont {Miyamoto},
  \citenamefont {Kuroda}, \citenamefont {Sekiyama}, \citenamefont {Yamagichi},
  \citenamefont {Fujiwara}, \citenamefont {Irizawa}, \citenamefont {Ito},
  \citenamefont {Kimura}, \citenamefont {Balashov}, \citenamefont {Wulfhekel},
  \citenamefont {Yeo}, \citenamefont {Iga},\ and\ \citenamefont
  {Imada}}]{Suga_2014}%
  \BibitemOpen
  \bibfield  {author} {\bibinfo {author} {\bibfnamefont {S.}~\bibnamefont
  {Suga}}, \bibinfo {author} {\bibfnamefont {K.}~\bibnamefont {Sakamoto}},
  \bibinfo {author} {\bibfnamefont {T.}~\bibnamefont {Okuda}}, \bibinfo
  {author} {\bibfnamefont {K.}~\bibnamefont {Miyamoto}}, \bibinfo {author}
  {\bibfnamefont {K.}~\bibnamefont {Kuroda}}, \bibinfo {author} {\bibfnamefont
  {A.}~\bibnamefont {Sekiyama}}, \bibinfo {author} {\bibfnamefont
  {J.}~\bibnamefont {Yamagichi}}, \bibinfo {author} {\bibfnamefont
  {H.}~\bibnamefont {Fujiwara}}, \bibinfo {author} {\bibfnamefont
  {A.}~\bibnamefont {Irizawa}}, \bibinfo {author} {\bibfnamefont
  {T.}~\bibnamefont {Ito}}, \bibinfo {author} {\bibfnamefont {S.}~\bibnamefont
  {Kimura}}, \bibinfo {author} {\bibfnamefont {T.}~\bibnamefont {Balashov}},
  \bibinfo {author} {\bibfnamefont {W.}~\bibnamefont {Wulfhekel}}, \bibinfo
  {author} {\bibfnamefont {S.}~\bibnamefont {Yeo}}, \bibinfo {author}
  {\bibfnamefont {F.}~\bibnamefont {Iga}}, \ and\ \bibinfo {author}
  {\bibfnamefont {S.}~\bibnamefont {Imada}},\ }\href@noop {} {\bibfield
  {journal} {\bibinfo  {journal} {J. Phys. Soc. Jpn.}\ }\textbf {\bibinfo
  {volume} {83}},\ \bibinfo {pages} {014705} (\bibinfo {year}
  {2014})}\BibitemShut {NoStop}%
\bibitem [{\citenamefont {Zhu}\ \emph {et~al.}(2013)\citenamefont {Zhu},
  \citenamefont {Nicolaou}, \citenamefont {Levy}, \citenamefont {Butch},
  \citenamefont {Syers}, \citenamefont {Wang}, \citenamefont {Paglione},
  \citenamefont {Sawatzky}, \citenamefont {Elfimov},\ and\ \citenamefont
  {Damascelli}}]{Zhu_2013}%
  \BibitemOpen
  \bibfield  {author} {\bibinfo {author} {\bibfnamefont {Z.-H.}\ \bibnamefont
  {Zhu}}, \bibinfo {author} {\bibfnamefont {A.}~\bibnamefont {Nicolaou}},
  \bibinfo {author} {\bibfnamefont {G.}~\bibnamefont {Levy}}, \bibinfo {author}
  {\bibfnamefont {N.~P.}\ \bibnamefont {Butch}}, \bibinfo {author}
  {\bibfnamefont {P.}~\bibnamefont {Syers}}, \bibinfo {author} {\bibfnamefont
  {X.~F.}\ \bibnamefont {Wang}}, \bibinfo {author} {\bibfnamefont
  {J.}~\bibnamefont {Paglione}}, \bibinfo {author} {\bibfnamefont {G.~A.}\
  \bibnamefont {Sawatzky}}, \bibinfo {author} {\bibfnamefont {I.~S.}\
  \bibnamefont {Elfimov}}, \ and\ \bibinfo {author} {\bibfnamefont
  {A.}~\bibnamefont {Damascelli}},\ }\href@noop {} {\bibfield  {journal}
  {\bibinfo  {journal} {Phys. Rev. Lett.}\ }\textbf {\bibinfo {volume} {111}},\
  \bibinfo {pages} {216402} (\bibinfo {year} {2013})}\BibitemShut {NoStop}%
\bibitem [{\citenamefont {Li}\ \emph {et~al.}(2014)\citenamefont {Li},
  \citenamefont {Li}, \citenamefont {Blaha},\ and\ \citenamefont
  {Kioussis}}]{Li_2014}%
  \BibitemOpen
  \bibfield  {author} {\bibinfo {author} {\bibfnamefont {Z.}~\bibnamefont
  {Li}}, \bibinfo {author} {\bibfnamefont {J.}~\bibnamefont {Li}}, \bibinfo
  {author} {\bibfnamefont {P.}~\bibnamefont {Blaha}}, \ and\ \bibinfo {author}
  {\bibfnamefont {N.}~\bibnamefont {Kioussis}},\ }\href@noop {} {\bibfield
  {journal} {\bibinfo  {journal} {Phys. Rev. B}\ }\textbf {\bibinfo {volume}
  {89}},\ \bibinfo {pages} {121117(R)} (\bibinfo {year} {2014})}\BibitemShut
  {NoStop}%
\bibitem [{\citenamefont {Herber}(2009)}]{Herber_2009}%
  \BibitemOpen
  \bibfield  {author} {\bibinfo {author} {\bibfnamefont {J.}~\bibnamefont
  {Herber}},\ }\href@noop {} {\bibfield  {journal} {\bibinfo  {journal}
  {Nature}\ }\textbf {\bibinfo {volume} {459}},\ \bibinfo {pages} {28}
  (\bibinfo {year} {2009})}\BibitemShut {NoStop}%
\bibitem [{\citenamefont {Mannhart}\ and\ \citenamefont
  {Schlom}(2010)}]{Mannhart_2010}%
  \BibitemOpen
  \bibfield  {author} {\bibinfo {author} {\bibfnamefont {J.}~\bibnamefont
  {Mannhart}}\ and\ \bibinfo {author} {\bibfnamefont {D.~G.}\ \bibnamefont
  {Schlom}},\ }\href@noop {} {\bibfield  {journal} {\bibinfo  {journal}
  {Science}\ }\textbf {\bibinfo {volume} {327}},\ \bibinfo {pages} {1607}
  (\bibinfo {year} {2010})}\BibitemShut {NoStop}%
\bibitem [{\citenamefont {Ji}\ \emph {et~al.}(2012)\citenamefont {Ji},
  \citenamefont {Allred}, \citenamefont {Ni}, \citenamefont {Tao},
  \citenamefont {Neupane}, \citenamefont {Wray}, \citenamefont {Xu},
  \citenamefont {Hasan},\ and\ \citenamefont {Cava}}]{Ji_2012}%
  \BibitemOpen
  \bibfield  {author} {\bibinfo {author} {\bibfnamefont {H.}~\bibnamefont
  {Ji}}, \bibinfo {author} {\bibfnamefont {J.~M.}\ \bibnamefont {Allred}},
  \bibinfo {author} {\bibfnamefont {N.}~\bibnamefont {Ni}}, \bibinfo {author}
  {\bibfnamefont {J.}~\bibnamefont {Tao}}, \bibinfo {author} {\bibfnamefont
  {M.}~\bibnamefont {Neupane}}, \bibinfo {author} {\bibfnamefont
  {A.}~\bibnamefont {Wray}}, \bibinfo {author} {\bibfnamefont {S.}~\bibnamefont
  {Xu}}, \bibinfo {author} {\bibfnamefont {M.~Z.}\ \bibnamefont {Hasan}}, \
  and\ \bibinfo {author} {\bibfnamefont {R.~J.}\ \bibnamefont {Cava}},\
  }\href@noop {} {\bibfield  {journal} {\bibinfo  {journal} {Phys. Rev. B}\
  }\textbf {\bibinfo {volume} {85}},\ \bibinfo {pages} {165313} (\bibinfo
  {year} {2012})}\BibitemShut {NoStop}%
\bibitem [{\citenamefont {Chang}\ and\ \citenamefont {{\it et.
  al.}}(2013)}]{Chang_2013}%
  \BibitemOpen
  \bibfield  {author} {\bibinfo {author} {\bibfnamefont {C.}~\bibnamefont
  {Chang}}\ and\ \bibinfo {author} {\bibnamefont {{\it et. al.}}},\ }\href@noop
  {} {\bibfield  {journal} {\bibinfo  {journal} {Science}\ }\textbf {\bibinfo
  {volume} {340}},\ \bibinfo {pages} {167} (\bibinfo {year}
  {2013})}\BibitemShut {NoStop}%
\bibitem [{\citenamefont {Ke}\ \emph {et~al.}(2013)\citenamefont {Ke},
  \citenamefont {Xu-Cun}, \citenamefont {Xi}, \citenamefont {Li}, \citenamefont
  {Ya-Yu},\ and\ \citenamefont {Qi-Kun}}]{Ke_2013}%
  \BibitemOpen
  \bibfield  {author} {\bibinfo {author} {\bibfnamefont {H.}~\bibnamefont
  {Ke}}, \bibinfo {author} {\bibfnamefont {M.}~\bibnamefont {Xu-Cun}}, \bibinfo
  {author} {\bibfnamefont {C.}~\bibnamefont {Xi}}, \bibinfo {author}
  {\bibfnamefont {L.}~\bibnamefont {Li}}, \bibinfo {author} {\bibfnamefont
  {W.}~\bibnamefont {Ya-Yu}}, \ and\ \bibinfo {author} {\bibfnamefont
  {X.}~\bibnamefont {Qi-Kun}},\ }\href@noop {} {\bibfield  {journal} {\bibinfo
  {journal} {Chin. Phys. B}\ }\textbf {\bibinfo {volume} {22}},\ \bibinfo
  {pages} {067305} (\bibinfo {year} {2013})}\BibitemShut {NoStop}%
\bibitem [{\citenamefont {Fu}\ and\ \citenamefont {Kane}(2009)}]{Fu_2009}%
  \BibitemOpen
  \bibfield  {author} {\bibinfo {author} {\bibfnamefont {L.}~\bibnamefont
  {Fu}}\ and\ \bibinfo {author} {\bibfnamefont {C.~L.}\ \bibnamefont {Kane}},\
  }\href@noop {} {\bibfield  {journal} {\bibinfo  {journal} {Phys. Rev. Lett.}\
  }\textbf {\bibinfo {volume} {102}},\ \bibinfo {pages} {216403} (\bibinfo
  {year} {2009})}\BibitemShut {NoStop}%
\bibitem [{\citenamefont {Akhmerov}\ \emph {et~al.}(2009)\citenamefont
  {Akhmerov}, \citenamefont {Nilsson},\ and\ \citenamefont
  {Beenakker}}]{Akhmerov_2009}%
  \BibitemOpen
  \bibfield  {author} {\bibinfo {author} {\bibfnamefont {A.~R.}\ \bibnamefont
  {Akhmerov}}, \bibinfo {author} {\bibfnamefont {J.}~\bibnamefont {Nilsson}}, \
  and\ \bibinfo {author} {\bibfnamefont {C.~W.~J.}\ \bibnamefont {Beenakker}},\
  }\href@noop {} {\bibfield  {journal} {\bibinfo  {journal} {Phys. Rev. Lett.}\
  }\textbf {\bibinfo {volume} {102}},\ \bibinfo {pages} {216404} (\bibinfo
  {year} {2009})}\BibitemShut {NoStop}%
\bibitem [{\citenamefont {Cook}\ and\ \citenamefont
  {Paramekanti}(2014)}]{Cook_2014}%
  \BibitemOpen
  \bibfield  {author} {\bibinfo {author} {\bibfnamefont {A.~M.}\ \bibnamefont
  {Cook}}\ and\ \bibinfo {author} {\bibfnamefont {A.}~\bibnamefont
  {Paramekanti}},\ }\href@noop {} {\bibfield  {journal} {\bibinfo  {journal}
  {Phys. Rev. Lett.}\ }\textbf {\bibinfo {volume} {113}},\ \bibinfo {pages}
  {077203} (\bibinfo {year} {2014})}\BibitemShut {NoStop}%
\bibitem [{\citenamefont {Cai}\ and\ \citenamefont {{\it et.
  al.}}(2013)}]{Cai_2013}%
  \BibitemOpen
  \bibfield  {author} {\bibinfo {author} {\bibfnamefont {T.}~\bibnamefont
  {Cai}}\ and\ \bibinfo {author} {\bibnamefont {{\it et. al.}}},\ }\href@noop
  {} {\bibfield  {journal} {\bibinfo  {journal} {arXiv:1310.2471}\ } (\bibinfo
  {year} {2013})}\BibitemShut {NoStop}%
\bibitem [{\citenamefont {Garrity}\ and\ \citenamefont
  {Vanderbilt}(2013)}]{Garrity_2013}%
  \BibitemOpen
  \bibfield  {author} {\bibinfo {author} {\bibfnamefont {K.~F.}\ \bibnamefont
  {Garrity}}\ and\ \bibinfo {author} {\bibfnamefont {D.}~\bibnamefont
  {Vanderbilt}},\ }\href@noop {} {\bibfield  {journal} {\bibinfo  {journal}
  {Phys. Rev. Lett.}\ }\textbf {\bibinfo {volume} {110}},\ \bibinfo {pages}
  {116802} (\bibinfo {year} {2013})}\BibitemShut {NoStop}%
\bibitem [{\citenamefont {Matthias}\ \emph {et~al.}(1961)\citenamefont
  {Matthias}, \citenamefont {Bozorth},\ and\ \citenamefont
  {Van~Vleck}}]{Matthias_1961}%
  \BibitemOpen
  \bibfield  {author} {\bibinfo {author} {\bibfnamefont {B.~T.}\ \bibnamefont
  {Matthias}}, \bibinfo {author} {\bibfnamefont {R.~M.}\ \bibnamefont
  {Bozorth}}, \ and\ \bibinfo {author} {\bibfnamefont {J.~H.}\ \bibnamefont
  {Van~Vleck}},\ }\href@noop {} {\bibfield  {journal} {\bibinfo  {journal}
  {Phys. Rev. Lett.}\ }\textbf {\bibinfo {volume} {7}},\ \bibinfo {pages} {160}
  (\bibinfo {year} {1961})}\BibitemShut {NoStop}%
\bibitem [{\citenamefont {Nolting}(1979)}]{Nolting_1979}%
  \BibitemOpen
  \bibfield  {author} {\bibinfo {author} {\bibfnamefont {W.}~\bibnamefont
  {Nolting}},\ }\href@noop {} {\bibfield  {journal} {\bibinfo  {journal} {Phys.
  Stat. Sol. (B)}\ }\textbf {\bibinfo {volume} {96}},\ \bibinfo {pages} {11}
  (\bibinfo {year} {1979})}\BibitemShut {NoStop}%
\bibitem [{\citenamefont {Adachi}\ and\ \citenamefont
  {Imanaka}(1998)}]{Adachi_1998}%
  \BibitemOpen
  \bibfield  {author} {\bibinfo {author} {\bibfnamefont {G.}~\bibnamefont
  {Adachi}}\ and\ \bibinfo {author} {\bibfnamefont {N.}~\bibnamefont
  {Imanaka}},\ }\href@noop {} {\bibfield  {journal} {\bibinfo  {journal} {Chem.
  Rev.}\ }\textbf {\bibinfo {volume} {98}},\ \bibinfo {pages} {1479} (\bibinfo
  {year} {1998})}\BibitemShut {NoStop}%
\bibitem [{\citenamefont {Steeneken}\ \emph {et~al.}(2002)\citenamefont
  {Steeneken}, \citenamefont {Tjeng}, \citenamefont {Elfimov}, \citenamefont
  {Sawatzky}, \citenamefont {Ghiringhelli}, \citenamefont {Brookes},\ and\
  \citenamefont {Huang}}]{Steeneken_2002}%
  \BibitemOpen
  \bibfield  {author} {\bibinfo {author} {\bibfnamefont {P.~G.}\ \bibnamefont
  {Steeneken}}, \bibinfo {author} {\bibfnamefont {L.~H.}\ \bibnamefont
  {Tjeng}}, \bibinfo {author} {\bibfnamefont {I.}~\bibnamefont {Elfimov}},
  \bibinfo {author} {\bibfnamefont {G.~A.}\ \bibnamefont {Sawatzky}}, \bibinfo
  {author} {\bibfnamefont {G.}~\bibnamefont {Ghiringhelli}}, \bibinfo {author}
  {\bibfnamefont {N.~B.}\ \bibnamefont {Brookes}}, \ and\ \bibinfo {author}
  {\bibfnamefont {D.~J.}\ \bibnamefont {Huang}},\ }\href@noop {} {\bibfield
  {journal} {\bibinfo  {journal} {Phys. Rev. Lett.}\ }\textbf {\bibinfo
  {volume} {88}},\ \bibinfo {pages} {047201} (\bibinfo {year}
  {2002})}\BibitemShut {NoStop}%
\bibitem [{\citenamefont {Schmel}\ \emph {et~al.}(2007)\citenamefont {Schmel},
  \citenamefont {Vaithyanathan}, \citenamefont {Herrnberger}, \citenamefont
  {Thiel}, \citenamefont {Richter}, \citenamefont {Liberati}, \citenamefont
  {Heeg}, \citenamefont {R{\"{o}}ckerath}, \citenamefont {Kourkoutis},
  \citenamefont {M{\"{u}}hlbauer}, \citenamefont {B{\"{o}}ni}, \citenamefont
  {Muller}, \citenamefont {Barash}, \citenamefont {Schubert}, \citenamefont
  {Idzerda}, \citenamefont {Mannhart},\ and\ \citenamefont
  {Schlom}}]{Schmehl_2007}%
  \BibitemOpen
  \bibfield  {author} {\bibinfo {author} {\bibfnamefont {A.}~\bibnamefont
  {Schmel}}, \bibinfo {author} {\bibfnamefont {V.}~\bibnamefont
  {Vaithyanathan}}, \bibinfo {author} {\bibfnamefont {A.}~\bibnamefont
  {Herrnberger}}, \bibinfo {author} {\bibfnamefont {S.}~\bibnamefont {Thiel}},
  \bibinfo {author} {\bibfnamefont {C.}~\bibnamefont {Richter}}, \bibinfo
  {author} {\bibfnamefont {M.}~\bibnamefont {Liberati}}, \bibinfo {author}
  {\bibfnamefont {T.}~\bibnamefont {Heeg}}, \bibinfo {author} {\bibfnamefont
  {M.}~\bibnamefont {R{\"{o}}ckerath}}, \bibinfo {author} {\bibfnamefont
  {L.}~\bibnamefont {Kourkoutis}}, \bibinfo {author} {\bibfnamefont
  {S.}~\bibnamefont {M{\"{u}}hlbauer}}, \bibinfo {author} {\bibfnamefont
  {P.}~\bibnamefont {B{\"{o}}ni}}, \bibinfo {author} {\bibfnamefont {D.~A.}\
  \bibnamefont {Muller}}, \bibinfo {author} {\bibfnamefont {Y.}~\bibnamefont
  {Barash}}, \bibinfo {author} {\bibfnamefont {J.}~\bibnamefont {Schubert}},
  \bibinfo {author} {\bibfnamefont {Y.}~\bibnamefont {Idzerda}}, \bibinfo
  {author} {\bibfnamefont {J.}~\bibnamefont {Mannhart}}, \ and\ \bibinfo
  {author} {\bibfnamefont {D.}~\bibnamefont {Schlom}},\ }\href@noop {}
  {\bibfield  {journal} {\bibinfo  {journal} {Nat. Mater.}\ }\textbf {\bibinfo
  {volume} {6}},\ \bibinfo {pages} {882} (\bibinfo {year} {2007})}\BibitemShut
  {NoStop}%
\bibitem [{\citenamefont {Sutarto}\ \emph {et~al.}(2009)\citenamefont
  {Sutarto}, \citenamefont {Altendorf}, \citenamefont {Coloru}, \citenamefont
  {Moretti~Sala}, \citenamefont {Haupricht}, \citenamefont {Chang},
  \citenamefont {Hu}, \citenamefont {Sch{\"{u}}ssler-Langeheine}, \citenamefont
  {Hollmann}, \citenamefont {Kierspel}, \citenamefont {Hsieh}, \citenamefont
  {J.}, \citenamefont {Chen},\ and\ \citenamefont {Tjeng}}]{Sutarto_2009}%
  \BibitemOpen
  \bibfield  {author} {\bibinfo {author} {\bibfnamefont {R.}~\bibnamefont
  {Sutarto}}, \bibinfo {author} {\bibfnamefont {S.~G.}\ \bibnamefont
  {Altendorf}}, \bibinfo {author} {\bibfnamefont {B.}~\bibnamefont {Coloru}},
  \bibinfo {author} {\bibfnamefont {M.}~\bibnamefont {Moretti~Sala}}, \bibinfo
  {author} {\bibfnamefont {T.}~\bibnamefont {Haupricht}}, \bibinfo {author}
  {\bibfnamefont {C.~F.}\ \bibnamefont {Chang}}, \bibinfo {author}
  {\bibfnamefont {Z.}~\bibnamefont {Hu}}, \bibinfo {author} {\bibfnamefont
  {C.}~\bibnamefont {Sch{\"{u}}ssler-Langeheine}}, \bibinfo {author}
  {\bibfnamefont {N.}~\bibnamefont {Hollmann}}, \bibinfo {author}
  {\bibfnamefont {H.}~\bibnamefont {Kierspel}}, \bibinfo {author}
  {\bibfnamefont {H.~H.}\ \bibnamefont {Hsieh}}, \bibinfo {author}
  {\bibfnamefont {L.~H.}\ \bibnamefont {J.}}, \bibinfo {author} {\bibfnamefont
  {C.~T.}\ \bibnamefont {Chen}}, \ and\ \bibinfo {author} {\bibfnamefont
  {L.~H.}\ \bibnamefont {Tjeng}},\ }\href@noop {} {\bibfield  {journal}
  {\bibinfo  {journal} {Phys. Rev. B}\ }\textbf {\bibinfo {volume} {79}},\
  \bibinfo {pages} {205318} (\bibinfo {year} {2009})}\BibitemShut {NoStop}%
\bibitem [{\citenamefont {Melville}\ \emph {et~al.}(2013)\citenamefont
  {Melville}, \citenamefont {Mairoser}, \citenamefont {Schmehl}, \citenamefont
  {Fischer}, \citenamefont {Gsell}, \citenamefont {Schreck}, \citenamefont
  {Awschalom}, \citenamefont {Heeg}, \citenamefont {Holl{\"{a}}nder},
  \citenamefont {Schubert},\ and\ \citenamefont {Schlom}}]{Melville_2013}%
  \BibitemOpen
  \bibfield  {author} {\bibinfo {author} {\bibfnamefont {A.}~\bibnamefont
  {Melville}}, \bibinfo {author} {\bibfnamefont {T.}~\bibnamefont {Mairoser}},
  \bibinfo {author} {\bibfnamefont {A.}~\bibnamefont {Schmehl}}, \bibinfo
  {author} {\bibfnamefont {M.}~\bibnamefont {Fischer}}, \bibinfo {author}
  {\bibfnamefont {S.}~\bibnamefont {Gsell}}, \bibinfo {author} {\bibfnamefont
  {M.}~\bibnamefont {Schreck}}, \bibinfo {author} {\bibfnamefont {D.~D.}\
  \bibnamefont {Awschalom}}, \bibinfo {author} {\bibfnamefont {T.}~\bibnamefont
  {Heeg}}, \bibinfo {author} {\bibfnamefont {B.}~\bibnamefont
  {Holl{\"{a}}nder}}, \bibinfo {author} {\bibfnamefont {J.}~\bibnamefont
  {Schubert}}, \ and\ \bibinfo {author} {\bibfnamefont {D.~G.}\ \bibnamefont
  {Schlom}},\ }\href@noop {} {\bibfield  {journal} {\bibinfo  {journal} {App.
  Phys. Lett.}\ }\textbf {\bibinfo {volume} {103}},\ \bibinfo {pages} {222402}
  (\bibinfo {year} {2013})}\BibitemShut {NoStop}%
\bibitem [{\citenamefont {Leger}\ \emph
  {et~al.}(1980{\natexlab{a}})\citenamefont {Leger}, \citenamefont {Yacoubi},\
  and\ \citenamefont {Loriers}}]{Leger_1980}%
  \BibitemOpen
  \bibfield  {author} {\bibinfo {author} {\bibfnamefont {J.~M.}\ \bibnamefont
  {Leger}}, \bibinfo {author} {\bibfnamefont {N.}~\bibnamefont {Yacoubi}}, \
  and\ \bibinfo {author} {\bibfnamefont {J.}~\bibnamefont {Loriers}},\
  }\href@noop {} {\bibfield  {journal} {\bibinfo  {journal} {Inorg. Chem.}\
  }\textbf {\bibinfo {volume} {19}},\ \bibinfo {pages} {2252} (\bibinfo {year}
  {1980}{\natexlab{a}})}\BibitemShut {NoStop}%
\bibitem [{\citenamefont {Krill}\ \emph {et~al.}(1980)\citenamefont {Krill},
  \citenamefont {Ravet}, \citenamefont {Kappler}, \citenamefont {Abadli},
  \citenamefont {Leger}, \citenamefont {Yacoubi},\ and\ \citenamefont
  {Loriers}}]{Krill_1980}%
  \BibitemOpen
  \bibfield  {author} {\bibinfo {author} {\bibfnamefont {G.}~\bibnamefont
  {Krill}}, \bibinfo {author} {\bibfnamefont {M.~F.}\ \bibnamefont {Ravet}},
  \bibinfo {author} {\bibfnamefont {J.~P.}\ \bibnamefont {Kappler}}, \bibinfo
  {author} {\bibfnamefont {L.}~\bibnamefont {Abadli}}, \bibinfo {author}
  {\bibfnamefont {J.~M.}\ \bibnamefont {Leger}}, \bibinfo {author}
  {\bibfnamefont {N.}~\bibnamefont {Yacoubi}}, \ and\ \bibinfo {author}
  {\bibfnamefont {C.}~\bibnamefont {Loriers}},\ }\href@noop {} {\bibfield
  {journal} {\bibinfo  {journal} {Solid State Comm.}\ }\textbf {\bibinfo
  {volume} {33}},\ \bibinfo {pages} {351} (\bibinfo {year} {1980})}\BibitemShut
  {NoStop}%
\bibitem [{\citenamefont {Leger}\ \emph
  {et~al.}(1980{\natexlab{b}})\citenamefont {Leger}, \citenamefont {Almonino},
  \citenamefont {Loriers}, \citenamefont {Dordor},\ and\ \citenamefont
  {Coqblin}}]{Leger_1980_a}%
  \BibitemOpen
  \bibfield  {author} {\bibinfo {author} {\bibfnamefont {J.~M.}\ \bibnamefont
  {Leger}}, \bibinfo {author} {\bibfnamefont {P.}~\bibnamefont {Almonino}},
  \bibinfo {author} {\bibfnamefont {J.}~\bibnamefont {Loriers}}, \bibinfo
  {author} {\bibfnamefont {P.}~\bibnamefont {Dordor}}, \ and\ \bibinfo {author}
  {\bibfnamefont {B.}~\bibnamefont {Coqblin}},\ }\href@noop {} {\bibfield
  {journal} {\bibinfo  {journal} {Phys. Lett.}\ }\textbf {\bibinfo {volume}
  {80A}},\ \bibinfo {pages} {325} (\bibinfo {year}
  {1980}{\natexlab{b}})}\BibitemShut {NoStop}%
\bibitem [{\citenamefont {Koepernik}\ and\ \citenamefont
  {Eschrig}(1999)}]{Koepernik_1999}%
  \BibitemOpen
  \bibfield  {author} {\bibinfo {author} {\bibfnamefont {K.}~\bibnamefont
  {Koepernik}}\ and\ \bibinfo {author} {\bibfnamefont {H.}~\bibnamefont
  {Eschrig}},\ }\href@noop {} {\bibfield  {journal} {\bibinfo  {journal} {Phys.
  Rev. B}\ }\textbf {\bibinfo {volume} {59}},\ \bibinfo {pages} {1743}
  (\bibinfo {year} {1999})}\BibitemShut {NoStop}%
\bibitem [{\citenamefont {Perdew}\ and\ \citenamefont
  {Wang}(1992)}]{Perdew_1992}%
  \BibitemOpen
  \bibfield  {author} {\bibinfo {author} {\bibfnamefont {J.~P.}\ \bibnamefont
  {Perdew}}\ and\ \bibinfo {author} {\bibfnamefont {Y.}~\bibnamefont {Wang}},\
  }\href@noop {} {\bibfield  {journal} {\bibinfo  {journal} {Phys. Rev. B}\
  }\textbf {\bibinfo {volume} {45}},\ \bibinfo {pages} {13244} (\bibinfo {year}
  {1992})}\BibitemShut {NoStop}%
\bibitem [{\citenamefont {Czy\.zyk}\ and\ \citenamefont
  {Swatzky}(1994)}]{Czyzyk_1994}%
  \BibitemOpen
  \bibfield  {author} {\bibinfo {author} {\bibfnamefont {M.~T.}\ \bibnamefont
  {Czy\.zyk}}\ and\ \bibinfo {author} {\bibfnamefont {G.~A.}\ \bibnamefont
  {Swatzky}},\ }\href@noop {} {\bibfield  {journal} {\bibinfo  {journal} {Phys.
  Rev. B}\ }\textbf {\bibinfo {volume} {49}},\ \bibinfo {pages} {14211}
  (\bibinfo {year} {1994})}\BibitemShut {NoStop}%
\bibitem [{\citenamefont {Tran}\ and\ \citenamefont {Blaha}(2009)}]{Tran_2009}%
  \BibitemOpen
  \bibfield  {author} {\bibinfo {author} {\bibfnamefont {F.}~\bibnamefont
  {Tran}}\ and\ \bibinfo {author} {\bibfnamefont {P.}~\bibnamefont {Blaha}},\
  }\href@noop {} {\bibfield  {journal} {\bibinfo  {journal} {Phys. Rev. Lett.}\
  }\textbf {\bibinfo {volume} {102}},\ \bibinfo {pages} {226401} (\bibinfo
  {year} {2009})}\BibitemShut {NoStop}%
\bibitem [{\citenamefont {Blaha}\ and\ \citenamefont
  {Schwarz}(2003)}]{Blaha_2003}%
  \BibitemOpen
  \bibfield  {author} {\bibinfo {author} {\bibfnamefont {P.}~\bibnamefont
  {Blaha}}\ and\ \bibinfo {author} {\bibfnamefont {K.}~\bibnamefont
  {Schwarz}},\ }\href@noop {} {\bibfield  {journal} {\bibinfo  {journal}
  {Comput. Mater. Sci.}\ }\textbf {\bibinfo {volume} {28}},\ \bibinfo {pages}
  {259} (\bibinfo {year} {2003})}\BibitemShut {NoStop}%
\bibitem [{\citenamefont {Martin}\ and\ \citenamefont
  {Allen}(1979)}]{Martin_1979}%
  \BibitemOpen
  \bibfield  {author} {\bibinfo {author} {\bibfnamefont {R.}~\bibnamefont
  {Martin}}\ and\ \bibinfo {author} {\bibfnamefont {J.}~\bibnamefont {Allen}},\
  }\href@noop {} {\bibfield  {journal} {\bibinfo  {journal} {J. Appl. Phys.}\
  }\textbf {\bibinfo {volume} {50}},\ \bibinfo {pages} {7561} (\bibinfo {year}
  {1979})}\BibitemShut {NoStop}%
\bibitem [{Note1()}]{Note1}%
  \BibitemOpen
  \bibinfo {note} {See Supplemental Material at
  http://link.aps.org/Supplemental/...}\BibitemShut {Stop}%
\bibitem [{\citenamefont {Zhang}\ and\ \citenamefont
  {Zhang}(2013)}]{Zhang_2013}%
  \BibitemOpen
  \bibfield  {author} {\bibinfo {author} {\bibfnamefont {H.}~\bibnamefont
  {Zhang}}\ and\ \bibinfo {author} {\bibfnamefont {S.}~\bibnamefont {Zhang}},\
  }\href@noop {} {\bibfield  {journal} {\bibinfo  {journal} {Phys. Status
  Solidi RRL}\ }\textbf {\bibinfo {volume} {7}},\ \bibinfo {pages} {72}
  (\bibinfo {year} {2013})}\BibitemShut {NoStop}%
\bibitem [{\citenamefont {Velev}\ and\ \citenamefont
  {Butler}(2007)}]{Velev_2007}%
  \BibitemOpen
  \bibfield  {author} {\bibinfo {author} {\bibfnamefont {J.}~\bibnamefont
  {Velev}}\ and\ \bibinfo {author} {\bibfnamefont {W.}~\bibnamefont {Butler}},\
  }\href@noop {} {\bibfield  {journal} {\bibinfo  {journal} {J. Phys.: Condens.
  Matter}\ }\textbf {\bibinfo {volume} {16}},\ \bibinfo {pages} {R637}
  (\bibinfo {year} {2007})}\BibitemShut {NoStop}%
\bibitem [{\citenamefont {L{\"{u}}thi}(1985)}]{Luthi_1985}%
  \BibitemOpen
  \bibfield  {author} {\bibinfo {author} {\bibfnamefont {B.}~\bibnamefont
  {L{\"{u}}thi}},\ }\href@noop {} {\bibfield  {journal} {\bibinfo  {journal}
  {J. Magn. Magn. Mat.}\ }\textbf {\bibinfo {volume} {52}},\ \bibinfo {pages}
  {70} (\bibinfo {year} {1985})}\BibitemShut {NoStop}%
\bibitem [{\citenamefont {Wachter}(1998)}]{Wachter_1994}%
  \BibitemOpen
  \bibfield  {author} {\bibinfo {author} {\bibfnamefont {P.}~\bibnamefont
  {Wachter}},\ }\href@noop {} {\emph {\bibinfo {title} {Handbook on the Physics
  and Chemistry of Rare Earths}}}\ (\bibinfo  {publisher} {Vol. 19, Chapter
  132},\ \bibinfo {year} {1998})\BibitemShut {NoStop}%
\end{thebibliography}

\begin{thebibliography}{56}%
\bibitem{Fu_2007} L. Fu, and C. L. Kane, Phys. Rev. B \textbf{76}, 045302 (2007).
\bibitem{Martin_1979} R. Martin, and J. Allen, J. Appl. Phys. \textbf{50}, 7561 (1979).
\end{thebibliography}


%

\newpage
\renewcommand{\thefigure}{S\arabic{figure}}
\renewcommand{\thetable}{S\arabic{table}}
\setcounter{figure}{0}

\begin{widetext}
\begin{center}
{\large\bf Supplemental Material}
\end{center}
\end{widetext}

\section{Calculation of topological indices}

\begin{figure}[b]
\centering
  \includegraphics[width=\columnwidth]{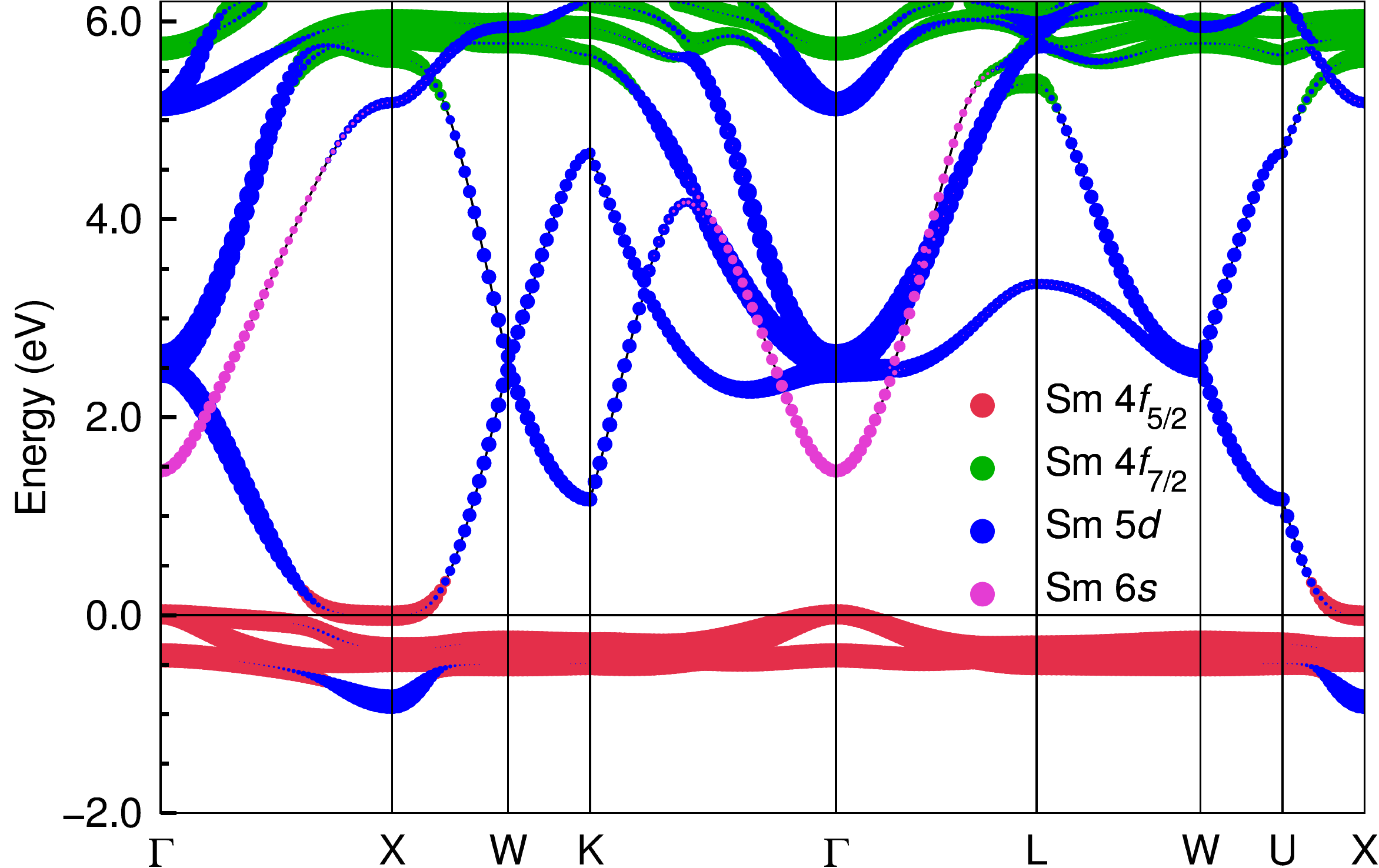}
  \includegraphics[width=\columnwidth]{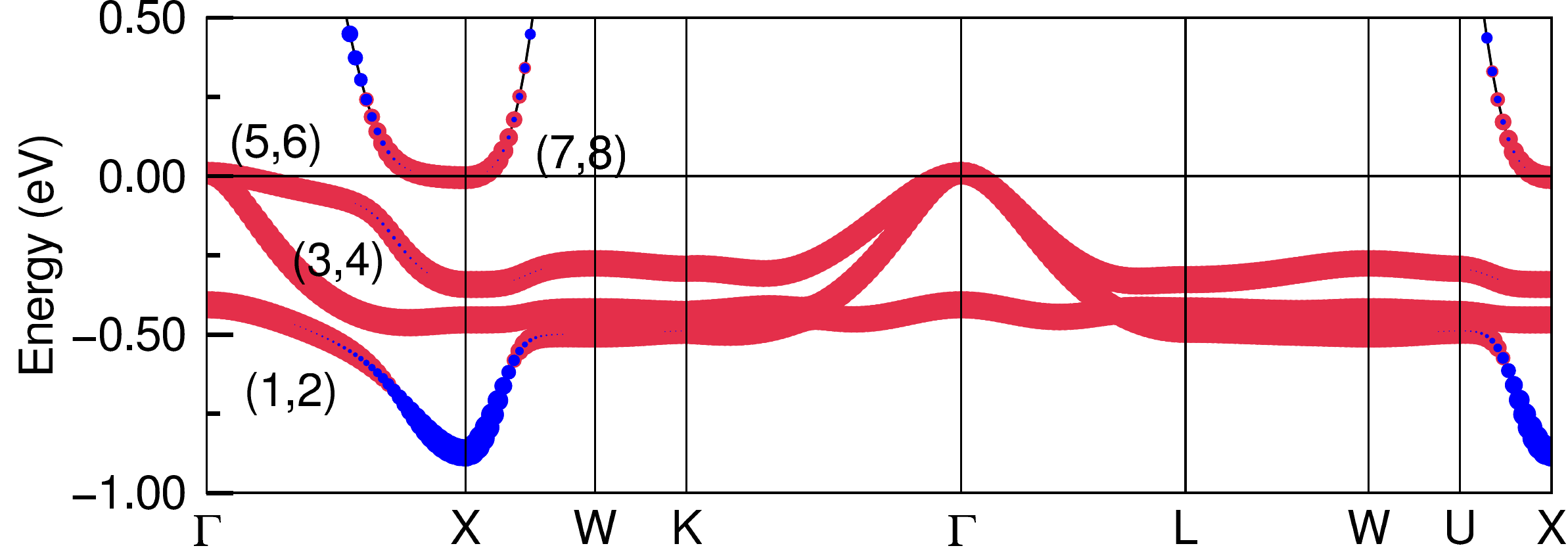}
\caption{(Color online) The calculated {\texttt{FPLO}} band structure of SmO ($a$ = 4.941\,\AA) within a non-spin polarized, full-relativistic 
approximation including strong Coulomb correlation 
(LDA+SO+$U$). A $U$  and $J_{H}$ value of 
6 eV and 0 eV have been used respectively. The size of the symbols represent the weight of the various orbital contributions to the underlying
bands. The numbers within parentheses are band indices to assist in assigning the various topological indices
in Table.\,\ref{tabnus}. }
\label{bands}
\end{figure}

To calculate the $\mathbb{Z}_{\mathrm{2}}$ topological index for a 
system with inversion symmetry, we use
the parity criteria as proposed by Fu and Kane\cite{Fu_2007}, wherein the product of
the parities of all occupied Kramers doublets at each time reversal
invariant momentum (TRIM) is determined. For a $fcc$
lattice, and hence a $bcc$ Brillouin zone, we have the following 8
TRIM points : 1 $\Gamma$ (0,0,0); 3 X (1,0,0)2$\pi/a$ and 4 L
($\frac{1}{2}$,$\frac{1}{2}$,$\frac{1}{2}$)2$\pi/a$.  The sign of the
product of the parities of all occupied doublets (Kramers degenerate
bands) at these eight TRIM points refer to a trivial topology, when
positive and to a nontrivial topology, when negative.  Under spatial
inversion, the $d$ orbitals are even while the $f$ orbitals are odd.
From the calculated band structures, it is clear that the occupied
bands between 0 to -2 eV at $\Gamma$ and the four L points possess $f$
orbital character, while due to the 4$f$-5$d$ band inversion, one of
the occupied band at the three X points possesses $d$ orbital
character.  The product of the parities for all the occupied bands
will then be negative, giving rise to a nontrivial topology in the
band structure of SmO.  This result is valid for both LDA+SO and
LDA+SO+$U$ calculations. This qualitative discussion was confirmed by
explicitly calculating the space group representations of the
inversion operator of all bands at the TRIM points (Table\,\ref{tabnus}). This gives access
to the four topological indices $\nu_0;(\nu_1\nu_2\nu_3)$. The
resulting indices for all bands up to the highest occupied
4$f_{\mathrm{5/2}}$ bands are $1;(000)$, which makes SmO strongly
topological. As a result of the particular arrangement of the bands around
the E$_{\mathrm{F}}$ in SmO, which creates a semi-metallic ``warped gap'' (no band
crossings at E$_{\mathrm{F}}$), it is
possible to apply the parity counting scheme at the TRIM points, since
this ``warped gap'' ensures the continuity condition for the phases of the
wave functions in the whole BZ which leads to the counting scheme. In
consequence one could classify SmO as a 3D strongly topological
semi-metal.

\begin{table}\caption{Calculated topological indices of selected bands in SmO within
LDA+SO+$U$ approximation ($a$ = 4.941\,\AA). $\varepsilon_{n}$ refers of the band energy at $\Gamma$; $n$ refers 
to the band indices, shown in Fig.\,\ref{bands}. $\Gamma$, X and L are the TRIM points. $\nu_{0};\left(\nu_{1}\nu_{2}\nu_{3}\right)$ are the four topological indices. The parity of the Kramers
doublet at the various TRIM points are denoted by O (odd) and E (even). Additionally, the
dominating orbital character of the bands at the TRIM points are listed.}\label{tabnus}
\begin{tabular}{l|l|l|l|c}
$\varepsilon_{n}\left(\Gamma\right)$ eV, ($n$) & $\Gamma$ & 3$\times$X & 4$\times$L & $\nu_{0};\left(\nu_{1}\nu_{2}\nu_{3}\right)$\tabularnewline
\hline 
 -3.650& O (2$p$) & O (2$p$) & E (2$p$) & $0;\left(000\right)$\tabularnewline
-0.493  (1,2) & O (4$f_{5/2}$) & E (5$d$) & O (4$f_{5/2}$) & $1;\left(000\right)$\tabularnewline
+0.030 (3,4) & O (4$f_{5/2}$) & O (4$f_{5/2}$) & O (4$f_{5/2}$) & $1;\left(000\right)$\tabularnewline
+0.030 (5,6) & O (4$f_{5/2}$) & O (4$f_{5/2}$) & O (4$f_{5/2}$) & $1;\left(000\right)$\tabularnewline
+1.457 (7,8) & E (6$s$) & O (5$d$) & E (5$d$) & $0;\left(000\right)$\tabularnewline
\end{tabular}
\end{table}

The topological indices switch from trivial to nontrivial at the
bottom of the six 4$f_{\mathrm{5/2}}$ derived bands and switch back to
trivial above the band, which forms the electron pockets around the X
point, which in consequence tells us that topological surface bands
could bridge energy window from the lowest 4$f_{\mathrm{5/2}}$ band to
the lowest unoccupied 5$d$ band.
In Table \ref{tabnus}, we provide the details of the topological character
of the important bands. The energy separation of about 2 eV between the
O2$p$ bands and the lowest 4$f$ state ensures that the gap
between O2$p$ and Sm 4$f$ must be trivial due to the fact that SO
coupling is too small to create band inversions bridging this gap.
Hence the highest O2$p$ band forms the trivial base line for
discussing the topological properties of the remaining bands. In the
table, we list the energy of the bands at the $\Gamma$ point, the band
indices (as indicated in Fig.\,\ref{bands}), the parity of the
Kramers doublets (O: odd, E: even) together with the dominating
orbital character at the three symmetry distinguished classes of TRIM
points as well as the resulting topological indices obtained from all
bands including the ones mentioned in the first column. Note that,
bands numbered 3$\ldots$6 are trivial bands, and do not change the
topology.

\section{Band structure using modified Becke Johnson approach}

\begin{figure}[h]
 \includegraphics[clip,width=\columnwidth]{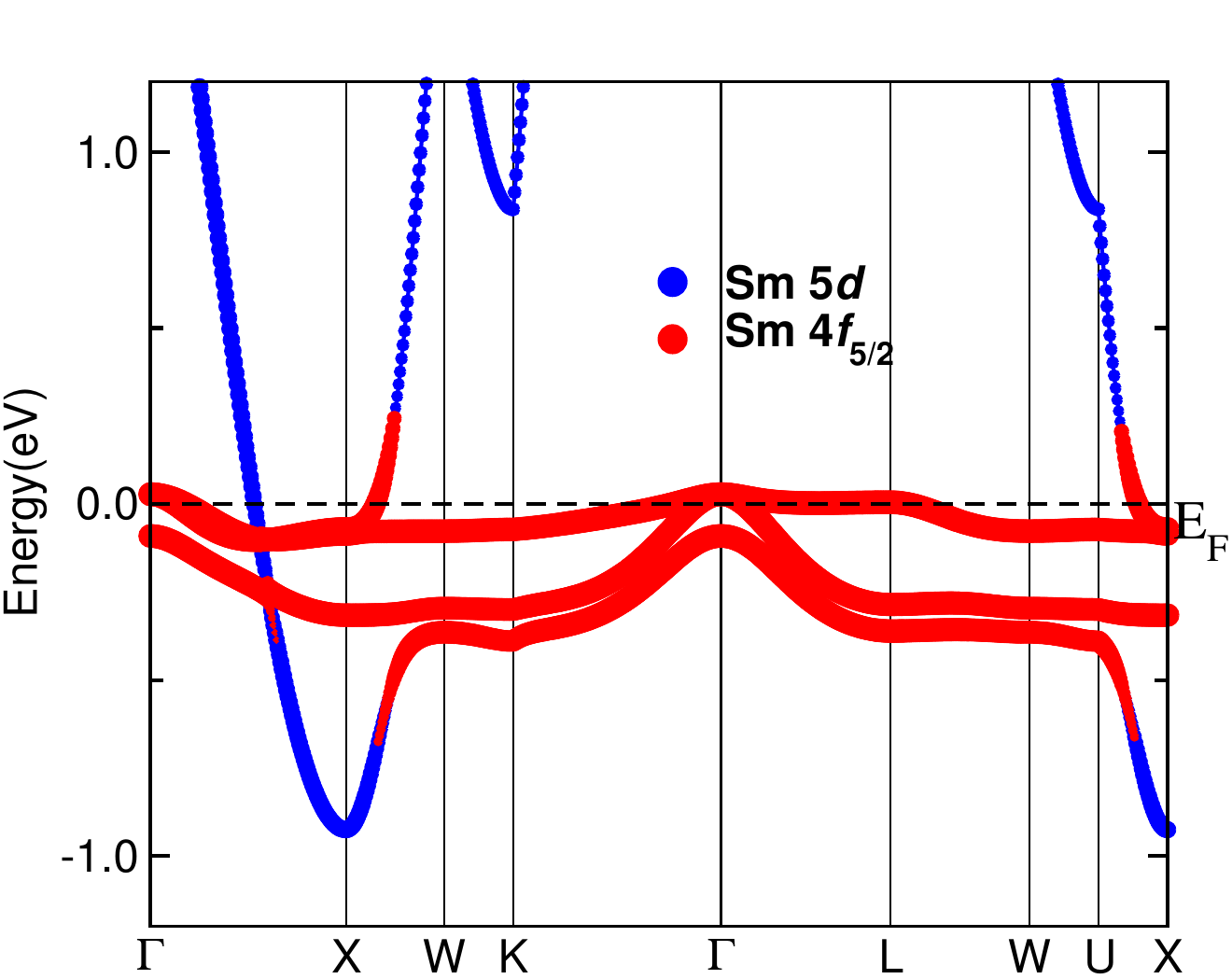}
  \includegraphics[clip,width=\columnwidth]{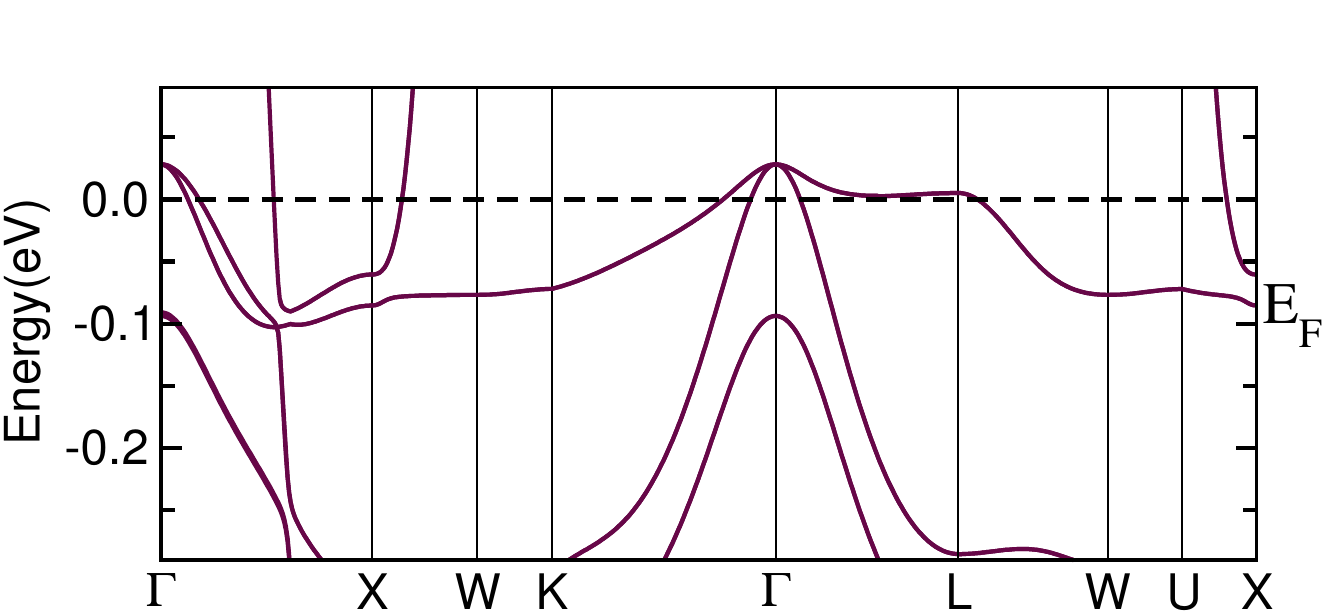}
\caption{(Color online) Band structure of SmO with orbital character using MBJLDA+SO approximation as implemented in {\texttt{WIEN2K}}. The topology and order of the bands are consistent to the ones obtained
using LDA+SO+$U$ scheme. The blow-up in the bottom panel clearly shows the opening of the warped band-gap along $\Gamma$-X, but smaller in size compared to {\texttt{FPLO}}. For clarity we have plotted the bands without any band character. }\label{wien2k}
\end{figure}

\begin{figure}[b]
 \includegraphics[clip,width=0.9\columnwidth]{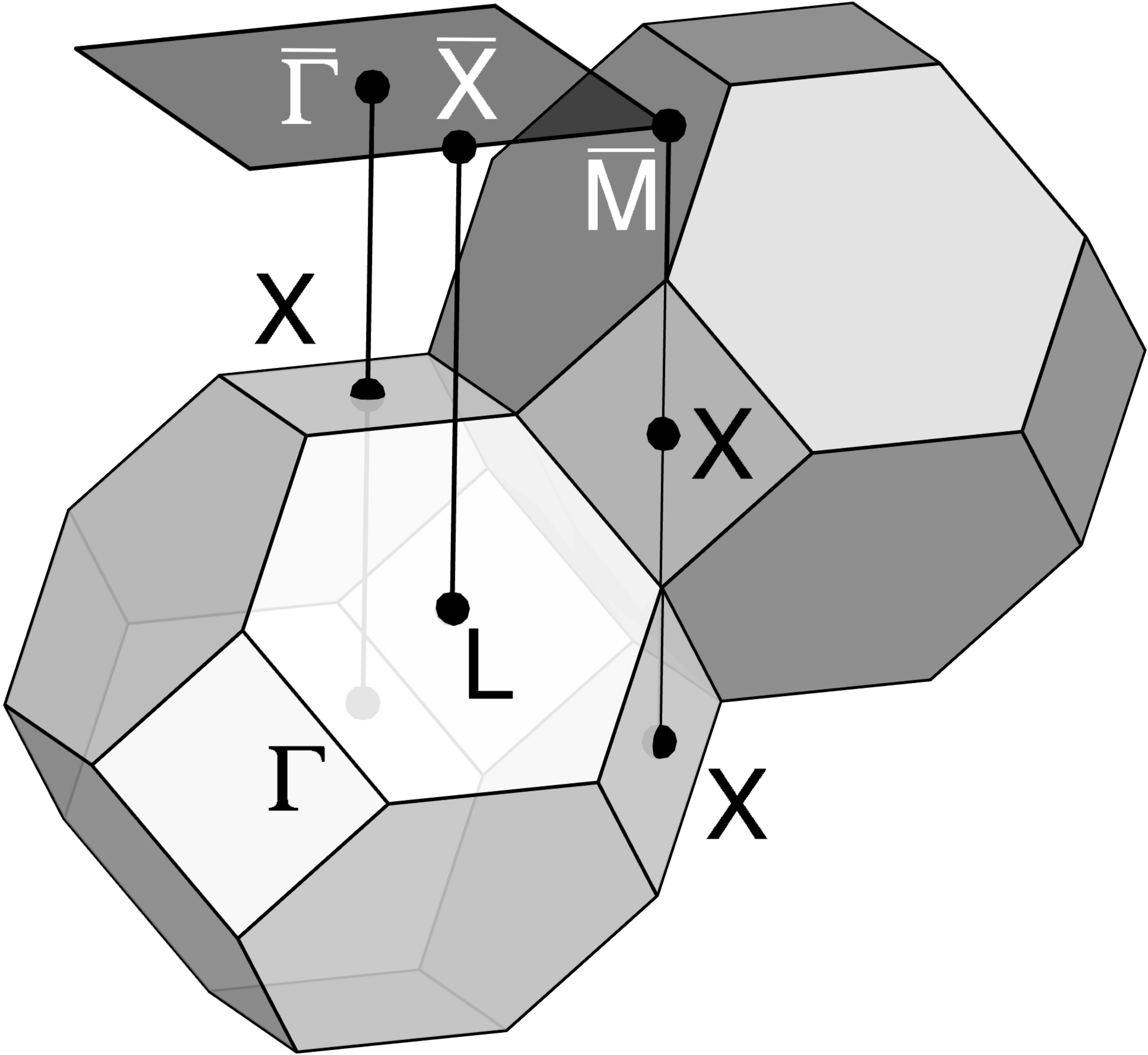}
\caption{(Color online) Bulk and surface Brillouin zone of a $fcc$ lattice. }\label{SBZ}
\end{figure}

To make sure that the topological characteristics
are not dependent on the choice of exchange and correlation functional, we calculated
the band structure of SmO using the recently proposed modified Becke-Johnson
functional (MBJLDA) as implemented in {\texttt{WIEN2K}}.
Fig.\,\ref{wien2k} shows the band structure obtained using MBJLDA+SO. 
The $f-d$ band inversion at $X$ point is consistent to our LDA+SO+$U$ calculations, though
the hybridization gap and the warped band-gap in MBJLDA+SO are smaller (but, of the same order).

\section{Surface state calculations}

To asses the nature of the surface band structure we calculated the Bloch spectral density 
($A_{\mathrm{Bl}}(k)$) of the 12 topmost surface layers of a semi infinite solid with [001] surface. 
For this purpose the DFT band structure was fitted with atom centered
maximally projected Wannier functions of Sm $5d$, $6s$ and $4f$ and  O $2p$ character. 
A soft energy cutoff was employed to achieve
band disentanglement at +6eV. The resulting fit models all relevant bands around the 
Fermi level within a few meV accuracy.
Afterwards, the obtained bulk hopping parameters were mapped onto a semi infinite solid 
and a Greens function technique was
used to calculate $A_{\mathrm{Bl}}(k)$ for the surface layers.
The resulting surface bands clearly show two weakly interacting Dirac cones around the surface projected $\bar{M}$-point.
In order to be of topology induced nature, there must be an odd number of Dirac cones. 
The surface Brillouin zone for a [001] surface
ends up with two bulk $X$-points being projected onto the surface $\bar{M}$-point and one 
bulk $X$-point being projected onto the surface $\bar{\Gamma}$-point (see Fig.\,\ref{SBZ}). Hence, one expects two 
Dirac cones at $\bar{M}$ and one at $\bar{\Gamma}$. The back-folding of the bulk band structure due to 
projection onto [001] leads to the appearance of electron pocket states from the bulk $X$-point at
$\bar{\Gamma}$, which overlap with the bulk projected states of the bulk $\Gamma$ hole pocket at $\bar{\Gamma}$. 
Hence, the third Dirac cone will be immersed into projected bulk states and can only form a surface resonance.
In order to prove the existence of this third Dirac cone, we performed an additional calculation, where the $4f$ electron pocket
around the bulk $\Gamma$-point are lowered in energy by application of a $\mathbf{k}$-dependent potential.
This removes bulk projected states from the low energy region at the surface
$\bar{\Gamma}$ point and reveals the third Dirac cone.
Finally, to rule out drastic effects of the relaxation of the surface electronic structure we performed a full DFT slab calculation
for 21 SmO layers. The resulting slab band structure shows the same two weakly interacting Dirac cones around the $\bar{M}$-point as the semi-infinite
calculation. Slight shifts of bands occur, but the overall structure equals that of the mapped model calculation.

\section{Notes on the many body state of SmO}
\label{app:MBstate}

The Sm $4f$ states are known to be strongly correlated. Nonetheless one can obtain useful information about this system using DFT within the local density approximation. The large onsite Coulomb repulsion between the $f$ electrons restricts the local valence occupations to fluctuate between the $f^5$ and $f^6$ configurations.
The lattice constants for the limiting cases of a pure $2+$ ($f^6$) and pure $3+$ ($f^5$) SmO compound are determined using spin polarized L(S)DA+SO+U. The local atomic $f^6$ ($f^5$) configuration has a lowest 
Hunds-rule multiplet state which belongs to the $^7F$ ($^6H$) term. LDA in the Kohn-Sham scheme does capture these local states as they are single Slater determinant representable (as any Hunds-rule high spin ground-state).

In order to determine the topology of the bands and in order to answer the question if hybridization between the Sm $d$ derived bands and the Sm $f$ derived bands is allowed, it is important to have the local symmetry correct. The spin polarized state found in L(S)DA does not represent the local Sm $f^6$ character correctly. Within L(S)DA the Sm $f^6$ configuration has a local moment ($M_z = 2 S_z + Lz = 3$) whereas the many body ground-state is a singlet ($\langle J^2 \rangle = J(J+1)=0$) without a local moment ($L_z=0$, $S_z=0$, $M_z=0$). The many-body ground-state without a local moment is consistent with experimental observations on divalent Sm compounds. In order to reproduce the local symmetry of the many body state of an $f^6$ configuration ($^7F_0$ term with $J=0$, belonging to the $A_{1g}$ representation) we use non-spin polarized DFT calculations. In this case the $f^{\mathrm{6}}$ configuration is represented by a state with six electrons in the $j=5/2$ bands. Although this state has an incorrect expectation value of $L$ ($\langle L^2 \rangle=24/7$) and $S$ ($\langle S^2 \rangle=24/7$) compared to the many body state (and thus a to high local Coulomb energy), it does have the right symmetry ($A_{1g}$) and total momentum ($J=0$). The $f^5$ configuration is represented by single hole excitations starting from the $f^6$ configuration. A peculiarity of starting from a many body $f^6$ state with $A_{1g}$ symmetry is that the many body one electron excitations have the same symmetry as the independent electron bands found in LDA, as shown by Martin and Allen\cite{Martin_1979}.

\end{document}